\documentclass[aps,prl,amsmath,amssymb,amsfont,showpacs,superscriptaddress,reprint]{revtex4-1}

\usepackage{graphicx}
\usepackage[usenames,dvips]{color}
\usepackage{times}
\usepackage{natbib}
\usepackage{subfigure}
\usepackage{soul}

\usepackage{xr}
\usepackage{cancel}
\usepackage{bm}

\begin{document}

\newcommand{\dd}[1]{\mathrm{d}#1}
\newcommand{\kb}{k_\text{B}}
\newcommand{\Td}{T_\text{D}}
\newcommand{\comment}[2]{#2}
\newcommand{\todo}[1]{{\color{red}#1}}

\newcommand{\vect}[1]{\bm{#1}}
\renewcommand{\Re}{ {\operatorname{Re}} }
\renewcommand{\Im}{ {\operatorname{Im}} }
\renewcommand{\d}{ {\mathrm{d}} } %for derivatives and integrals

\newcommand{\JO}[1]{\color{blue}#1}
\newcommand{\jav}[1]{\textcolor{red}{#1}}

%\title{Loschmidt echo and the unified theory of spin-orbit coupling induced spin-relaxation in solids}

%\title{Loschmidt echo and a concise formula of spin-orbit coupling induced spin-relaxation in solids}

%A concise formula of SOC induced spin relaxation in solid and the Loschmidt echo of a spin ensemble

%\title{The Loschmidt echo in spin dephasing and the generic expression of spin-orbit coupling induced spin-relaxation in solids}
%\title{A generic description of spin relaxation in solids and the Loschmidt echo}
%\title{The Loschmidt echo and a generic expression of spin-orbit coupling induced spin relaxation}
%\title{The generic phase diagram of spin relaxation in semiconductors and the Loschmidt echo}
\title{The generic phase diagram of spin relaxation in solids and the Loschmidt echo}

\author{G\'{a}bor Cs\H{o}sz}
\thanks{These two authors contributed equally}
\affiliation{Department of Physics, Budapest University of Technology and Economics and
MTA-BME Lend\"{u}let Spintronics Research Group (PROSPIN), POBox 91, H-1521 Budapest, Hungary}

\author{L\'en\'ard Szolnoki}
\thanks{These two authors contributed equally}
\affiliation{Department of Physics, Budapest University of Technology and Economics and
MTA-BME Lend\"{u}let Spintronics Research Group (PROSPIN), POBox 91, H-1521 Budapest, Hungary}

\author{Annam\'aria Kiss}
\affiliation{Institute for Solid State Physics and Optics, Wigner Research Centre for Physics, POBox 49, H-1525 Budapest, Hungary} %Hungarian Academy of Sciences, 
\affiliation{Department of Physics, Budapest University of Technology and Economics and
MTA-BME Lend\"{u}let Spintronics Research Group (PROSPIN), POBox 91, H-1521 Budapest, Hungary}

\author{Bal\'azs D\'ora}
\affiliation{Department of Theoretical Physics, Budapest University of Technology and Economics and
MTA-BME Lend\"{u}let Topology Research Group (TOPCOR), POBox 91, H-1521 Budapest, Hungary}

%\author{Jaroslav Fabian}
%\affiliation{Institute for Theoretical Physics, University of Regensburg, 93040 Regensburg, Germany}

\author{Ferenc Simon}
\email{f.simon@eik.bme.hu}
\affiliation{Department of Physics, Budapest University of Technology and Economics and
MTA-BME Lend\"{u}let Spintronics Research Group (PROSPIN), POBox 91, H-1521 Budapest, Hungary}

%76.30.Pk Electron paramagnetic resonance and relaxation Conduction electrons,(BULK)	 71.70.Ej Spin-orbit coupling, Zeeman and Stark splitting, Jahn-Teller effect, 72.25.Rb	 Spin relaxation and scattering 75.76.+j	Spin transport effects (for devices exploiting spin polarized transport, see 85.75.Hh, 85.75.Mm, and 85.75.Ss)

\date{\today}

\begin{abstract}
The spin relaxation time in solids is determined by several competing energy scales and processes and distinct methods
are called for to analyze the various regimes.
We present a stochastic model for the spin dynamics in solids which is equivalent to solving the spin Boltzmann equation and takes the relevant processes into account on equal footing.
The calculations reveal yet unknown parts of the spin-relaxation phase diagram, where strong spin-dephasing occurs in addition 
to spin-relaxation. Spin-relaxation times are obtained for this regime by introducing the numerical Loschmidt echo. 
This allowes us to construct a generic approximate formula for the spin-relaxation time, $\tau_{\text{s}}$, for the entire 
phase diagram, involving the quasiparticle scattering rate, $\Gamma$, spin-orbit coupling strength, $\mathcal{L}$, and a magnetic term, 
$\Delta_{\text{Z}}$ due to the Zeeman effect. The generic expression reads as $\hbar/\tau_{\text{s}}\approx \Gamma\cdot \mathcal{L}^2 /(\Gamma^2+\mathcal{L}^2+\Delta_{\text{Z}}^2)$.
\end{abstract}

\maketitle

%\section*{Introduction}
\textit{Introduction.} The emerging field of spintronics \cite{WolfSCI,FabianRMP} envisions to employ the electron spin as information carrier instead of the usual charge degree of freedom, thus allowing for more efficient and high performance future informatics devices. This potential lead to renewed efforts for both the theoretical understanding of spintronics relevant phenomena \cite{FabianRMP,WuReview} and also to explore novel materials for this purpose. In particular, two-dimensional materials, such as graphene \cite{NovoselovSCI2004} appear to be excellent spintronics candidates \cite{TombrosNat2007,KawakamiPRL2010,GuntherodtBilayer}. 

Whether a material can be successfully employed in spintronics is decided by the magnitude of the so-called spin-relaxation time, $\tau_{\text{s}}$, and the related spin-diffusion length, $\delta_{\text{s}}$. These are the analogous quantities to the charge carrier life-time and carrier diffusion-length in semiconductors. There is an overall consensus that spin-relaxation is dominated by spin-orbit coupling (SOC) effects \cite{FabianRMP,WuReview}. The Elliott-Yafet (EY) \cite{Elliott,yafet1963g} and D'yakonov-Perel' (DP) \cite{DyakonovPerelSPSS1972,PikusTitkov} theories explain spin-relaxation in metals and semiconductors with and without inversion symmetry (e.g. GaAs), respectively. Although these two descriptions are formulated differently, these were recently brought to a common mathematical basis \cite{SzolnokiPRB2017}.

Modern advances in the description of spin-relaxation aim at accurately determining $\tau_{\text{s}}$ with first-principles methods including the details of the crystal, band structure, and electron-phonon coupling \cite{Schneider,WuNJP2012,RestrepoWindl,WuYuPRB2016,Wu_TMDC}. However, it turned out recently that conventional spin-relaxation theories require refinement \cite{BorossSciRep2013,SzolnokiSciRep2017,SzolnokiPRB2017}, which also affects the first principles descriptions. A representative example is the case of graphite, where first principles prediction gives a temperature dependence of $\tau_{\text{s}}$ that is opposite to the experimentally observed one \cite{RestrepoWindl}.  

%\begin{figure*}[htp]
\begin{figure}[htp]
\begin{center}
\includegraphics[scale=1.1]{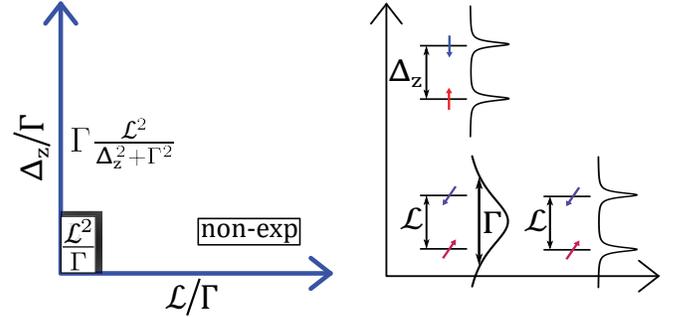}
\caption{\textit{Left panel:} Phase diagram of spin-relaxation for materials without inversion symmetry. The DP \cite{DyakonovPerelSPSS1972,PikusTitkov} regime is indicated by a shaded box in the bottom left corner. The other known result for the $\Gamma\--\Delta_{\text{Z}}$ line (Ref. \onlinecite{BorossSciRep2013}) is also given. The large $\mathcal{L}$ regime leads to a non-exponentially dephasing regime \cite{FabianRMP,SzolnokiSciRep2017}, which is tackled herein. \textit{Right panel:} Energy level schemes for the different relaxation regimes; the gaps, spin degeneracy, and the quasiparticle spectral function is indicated. Vertical and tilted arrows depict spin and SOC eigenstates, respectively.}
\label{fig:Artwork}
\end{center}
%\end{figure*}
\end{figure}

Spin relaxation time (and the correponding spin-relaxation rate, $\Gamma_{\text{s}}=\hbar/\tau_{\text{s}}$) is strongly influenced by the momentum scattering time, $\tau$ (or the strength of quasiparticle scattering, $\Gamma=\hbar/\tau$), by the magnitude of an external magnetic field and by the spin-orbit coupling. The magnetic field induced Zeeman splitting is characterized by $\Delta_{\text{Z}}$ and the SOC strength is characterized by its Fermi surface averaged effective matrix element, $\mathcal{L}$, but it can be associated with a built-in, SOC-related magnetic field. The different spin-relaxation regimes are summarized in Fig. \ref{fig:Artwork} along with the corresponding band structure.

%Note that we denote by $\Delta$ the gaps with two different origins: $\Delta$ in the EY theorem refers to the kinetic energy level spacing gap whereas the magnetic field plays the same role in materials without the inversion symmetry thus $\Delta$ denotes the sum of the spin and orbital Zeeman energy splitting.

The DP regime is highlighted in the figure, which occurs when $\Gamma\gg(\mathcal{L},\Delta_{\text{Z}})$. The behavior of spin relaxation was recently given for $\mathcal{L}\ll(\Gamma,\Delta_{\text{Z}})$ (Ref. \onlinecite{BorossSciRep2013}). The large $\mathcal{L}$ regime was described in Ref. \onlinecite{FabianRMP} to give rise to a strongly non-exponential spin decay or dephasing. Spin relaxation (SR) and dephasing have very different characteristics: SR is a truly irreversible process (in other words, it involves a memory loss and an entropy increase), whereas dephasing is at least partially reversible as it does not involve a full memory loss. The concept of the Loschmidt echo \cite{Loschmidt_Scholarpedia} was introduced for such situations in a famous \textit{Ge\-dan\-ken\-ex\-per\-i\-ment} to allow separation of the two effects. In practice, the most successful realization of the Loschmidt echo is the spin-echo \cite{Hahn1950}.

Description of spin-relaxation is lacking for the dephasing regime. This motivated us to develop a numerical approach to the dynamics of spins which includes momentum scattering and spin precession under the action of an external and the SOC related magnetic fields. The method provides the quantum trajectories \cite{QTrajectory1,QTrajectory2,QTrajectory3} for individual spins and is shown to be equivalent to the exact solution of the spin Boltzmann equation. Spin-relaxation time can be obtained even in the presence of strong dephasing with the introduction of a numerical 
Loschmidt echo. This allowed us to construct the full phase diagram of $\left(\Gamma,\mathcal{L},\Delta_{\text{Z}}\right)$ and we find that a generic formula: 
\begin{gather}
\Gamma_{\text{s}} \approx \Gamma \cdot \frac{\mathcal{L}^2}{\Gamma^2+\mathcal{L}^2+\Delta_{\text{Z}}^2}
\end{gather}
fits well the data for the \emph{entire} phase diagram.

\textit{The stochastic model.} Dynamics of electron spins in semiconductors is described by the time evolution of the density matrix of spins, 
$\rho_{\vect{k}}$. This leads to the so-called spin Boltzmann equation \cite{FabianActaPhysSlovaca}:

\begin{equation}
	\frac{\partial \rho_{\vect{k}}}{\partial t}-\frac{1}{i\hbar}\left[H_{\text{Z},\vect{k}},\rho_{\vect{k}}\right]=\sum_{\vect{k'}}W_{\vect{k}\vect{k'}}\left(\rho_{\vect{k'}}-\rho_{\vect{k}}\right),
\end{equation}
which takes the spin into account quantum mechanically, while the other degrees of freedom are treated quasiclassically.
The second term on the left side gives spin evolution under the action of the $H_{\text{Z},\vect{k}}$ Zeeman Hamiltonian which contains the vectorial sum of an external and the built-in SOC related (thus $k$-dependent) magnetic 
fields. $W_{\vect{k}\vect{k'}}$ is the $k\rightarrow k'$ scattering probability per unit time which obeys the detailed balance. The spin expectation value, $\vect{s}_{\vect{k}}=\mathrm{Tr}\left[\rho_{\vect{k}}\vect{\sigma}\right]$, reads:

\begin{equation}
	\frac{\partial\vect{s}_{\vect{k}}}{\partial t}=\vect{\Omega}\left(\vect{k}\right)\times\vect{s}_{\vect{k}}+\sum_{\vect{k'}}W_{\vect{k}\vect{k'}}\left(\vect{s}_{\vect{k'}}-\vect{s}_{\vect{k}}\right).
	\label{eq::spin_Boltzmann}
\end{equation}
The Larmor precession term is $H_{\text{Z},\vect{k}}=\frac{\hbar}{2}\cdot \vect{\Omega}\left(\vect{k}\right)\vect{\sigma}$. In principle, the numerical solution of Eq. \eqref{eq::spin_Boltzmann} could provide the full spin dynamics for various values of the momentum scattering rate, external magnetic field, and SOC strength. 

%\begin{figure*}[htp]
\begin{figure}[htp]
\begin{center}
\includegraphics[scale=.18]{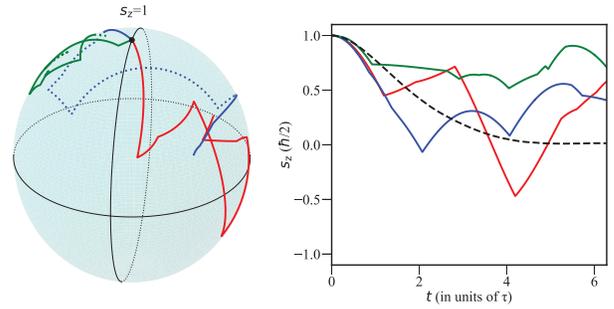}
\caption{Quantum trajectories of three individual spins on the Bloch sphere (dotted curves are trajectories on its back). All spins are polarized at $t=0$ to $s_{z}=1$, and they evolve under the action of the external and SOC magnetic field. Elapsed time between two momentum scatterings follows a Poisson distribution with expectation value of $\tau$. Time evolution of the $s_z$ component is shown on the right panel and dashed curve is the ensemble averaged time evolution of $s_z$.}
\label{fig:spin_trajectories}
\end{center}
%\end{figure*}
\end{figure}

Our stochastic or Monte Carlo (MC) model considers a spin ensemble with different $\mathbf{k}$'s where all spins are initially polarized along the $z$ axis on the Bloch sphere. The spins evolve independently and they undergo Larmor precession between two momentum scattering events. Momentum scattering randomizes $\mathbf{k}$ and precession continues with a new Larmor precession vector. The elapsed time between two momentum scatterings follows a Poisson distribution, with expectation value of $\tau_{\vect{k}}=\left(\sum_{\vect{k'}}W_{\vect{k}\rightarrow \vect{k'}}\right)^{-1}$. Fig. \ref{fig:spin_trajectories}. shows the quantum trajectories \cite{QTrajectory1,QTrajectory2,QTrajectory3} of three individual electron spins on the Bloch sphere along with the ensemble average value of $s_z$.

The time evolution of an individual spin in state $\vect{k}$ during a short $\Delta t$ time interval is:

\begin{equation}
\begin{aligned}
	\vect{s}_{\vect{k}}(t+\Delta t)=\left(1-\frac{\Delta t}{\tau_{\vect{k}}}\right)\vect{U}_{\vect{k}}(\Delta t)\vect{s}_{\vect{k}}(t)+\Delta t\sum_{\vect{k'}}W_{\vect{k'}\vect{k}}\vect{s}_{\vect{k'}}(t),
\end{aligned}
\end{equation}
where $\vect{U}_{\vect{k}}\left(\Delta t\right)\vect{s}_{\vect{k}}(t)=\vect{s}_{\vect{k}}(t)+\Delta t\vect{\Omega}\left(\vect{k}\right)\times \vect{s}_{\vect{k}}(t)$. The first term on the right hand side describes when the spin, $s_{\vect{k}}$ does not scatter out from $\vect{k}$ during $\Delta t$. The second corresponds to scattering-in from $\vect{k'}$ states. Assuming detailed balance ($W_{\vect{k'k}}=W_{\vect{kk'}}$) and retaining terms in first order of $\Delta t$ yields directly the spin Boltzmann equation of Eq. \eqref{eq::spin_Boltzmann} in the infinitesimal limit of the time step. We note that this consideration essentially mimics the derivation of the Lindblad equation in Ref. \onlinecite{MCWF3}.

Besides being equivalent, the MC method is numerically more effective as the Larmor precession between scattering events can be calculated analytically without resorting to numerics. In contrast, solving the spin Boltzmann equation for a large magnetic field (external or SOC related) requires the use of a small time discretization. 

The most important advantage of the MC method is that it tracks the invidual quantum trajectories of electron spins, whereas the spin Boltzmann equation inherently provides the ensemble average values only, similarly to the Schr\"odinger equation. The MC method is particularly advantageous when we are interested in single events and quantum leaps happening in individual quantum systems, e.g. in mesoscopic systems where statistical fluctuations are important in understanding and analyzing individual measurements \cite{QTrajectory3}. We employ the MC method in the following to study yet unexplored parts of the spin-relaxation phase diagram as a function of $\left(\Gamma,\mathcal{L},\Delta_{\text{Z}}\right)$. 

%The method considers a spin ensemble where all spins are initially polarized along the $z$ axis. The spins, which are considered as classical variables, Larmor precess under the action of the built-on magnetic field which is generated by the SOC. The Larmor precession vector is randomized after each momentum scattering, which occur with a mean time separation of $\tau$. The spin values are ensemble averaged and projected along the $z$ axis to yield the amount of spin direction decay. This is in fact the conventional description of the DP theory\cite{DyakonovPerelSPSS1972,PikusTitkov}, which can be intuitively adapted numerically in practice. The method was validated by comparing the simulated spin-relaxation times with values obtained from analytically solvable models \cite{SzolnokiSciRep2017}. 

\textit{Non-exponentional spin relaxation.} We apply the MC method for Dresselhaus SOC Hamiltonian in three dimensions, where the Larmor (angular) frequency vectors read in $k$-space ($\mathbf{k}=(k_x,k_y,k_z)$):
       \begin{equation}
        \mathbf{\Omega}(\mathbf{k})
        = \frac{\mathcal{L'}}{\hbar k^3_{\text{F}}}
          \left[
            k_x\left(k_y^2-k_z^2\right),
            k_y\left(k_z^2-k_x^2\right),
            k_z\left(k_x^2-k_y^2\right)
          \right],
      \end{equation}
where $\mathcal{L'}$ is the strength of the SOC in energy units, $k_{\text{F}}$ is the Fermi wavenumber. The Fermi surface averaging of $\mathcal{L'}$ gives $\mathcal{L}^2=\mathcal{L'}^2\cdot \frac{4}{35}$.

We first consider a zero magnetic field, i.e. $\Delta_{\text{Z}}=0$. The condition of the DP description \cite{DyakonovPerelSPSS1972,PikusTitkov} is $\Gamma \gg \mathcal{L}$, when the MC method gives spin relaxation with a single exponent as $\Gamma_{\text{s}}=\frac{\mathcal{L}^2}{\Gamma}$. However, the situation drastically changes when $\mathcal{L}$ is the leading term and a non-exponential spin decay is observed. This regime was first described in Ref. \onlinecite{FabianRMP}: it was pointed out that a significant dephasing, rather than relaxation, takes place on a timescale of $1/\Delta\Omega$, where the latter is the spread in the Larmor frequency distribution. The qualitative reason for dephasing is that the spins precess by a large angle, $\overline{\Omega}\tau \gg 1$ before a momentum scattering takes place ($\overline{\Omega}$ is the mean value of the Larmor frequency). The dephasing is in fact a procedure without memory loss, which is followed by a truly irreversible spin decay after several $\tau$ elapses \cite{ShermanPRB2011,ShermanPRB2018}.

\begin{figure}[htp]
\begin{center}
\includegraphics[scale=0.45]{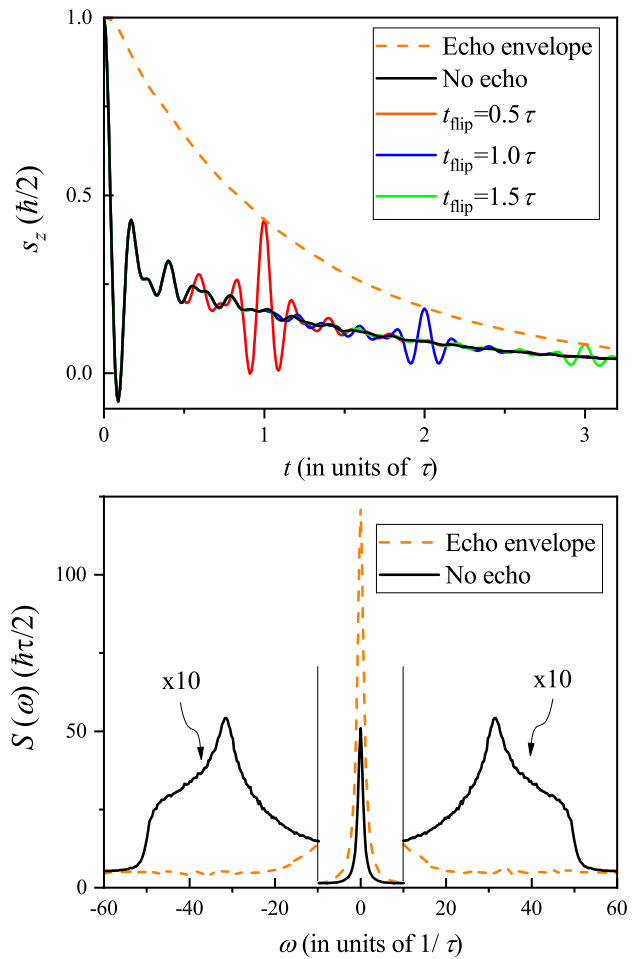}
\caption{\textit{Upper panel:} Spin decay for a spin ensemble which are spin polarized at $t=0$ (black solid line) and are under the action of a strong Dresselhaus SOC, which leads to dephasing. Loschmidt echoes (red, blue, and green lines) are generated by inverting the Larmor frequency vectors at $t=\tau/2,\,\tau$, and $1.5\tau$ and these are observed around $t=\tau,\,2\tau$, and $3\tau$, respectively. The envelope of these echoes is depicted with a dashed orange line. \textit{Lower panel:} Fourier transform of the spin decay and the Loschmidt echo envelopes. Note that the latter contains a single peak, centered at $\omega=0$, whereas the earlier has a peak at $\omega=0$ and two side-lobe structures (shown on a magnified scale), which reflect the Larmor frequency distribution in the Dresselhaus SOC.}
\label{fig:SpinDecay_Loschmidt}
\end{center}
\end{figure}

The corresponding spin decay is shown in Fig. \ref{fig:SpinDecay_Loschmidt} (solid black line in the upper panel). The simulation was performed for an ensemble of $N=10^{5}$ electrons and a nominal value of the Larmor precession frequency of $\hbar\Omega/\Gamma=\mathcal{L'}/\Gamma=100$. Clearly, the data shows a rapid dephasing, which is followed by an exponential-like tail for longer times. Its Fourier transform, $\mathcal{S}(\omega)$, is presented in the lower panel: it shows a peak function at $\omega=0$ and two side-lobe structures, which are due to the distribution of the Larmor frequencies in the Dresselhaus SOC.

The simultaneous presence of dephasing and relaxation is encountered in magnetic resonance and is tackled with the concept of spin echo \cite{SlichterBook,AbragamBook}, which is discussed in detail in the Supplementary Material \cite{SupMat}. The spin echo is a specific case of the Loschmidt echo \cite{Loschmidt_Scholarpedia}. In our case, the Loschmidt echo can be numerically introduced by \emph{inverting} the $\mathbf{\Omega}(\mathbf{k})$ vectors at a given instant and observing the recovery of the ensemble spin value.
Fig. \ref{fig:SpinDecay_Loschmidt}. shows three such echoes which were generated by inverting the $\mathbf{\Omega}(\mathbf{k})$ vectors at three different instances. The envelope of the echoes is also show in the figure. We developed a numerical method \cite{SupMat} to obtain the Loschmidt echo envelopes without needing to calculate the echoes at each time points. The corresponding Fourier transform, $\mathcal{S}(\omega)$, of the Loschmidt echo envelope is shown in the lower panel of Fig. \ref{fig:SpinDecay_Loschmidt}: as expected, it shows a single peak at $\omega=0$ which contains all the spectral weight from the two side-lobes. We highlight an interesting analogy of the present calculations with magnetic resonance: the timescale of spin dephasing corresponds to $T_2^*$ (often referred to as reversible "relaxation time") and the envelope to the $T_2$ spin-spin relaxation time (also known as irreversible relaxation time).

Inverting the SOC related Larmor frequency vectors is not practically conceivable in bulk solid state realizations. However, inverting an external electric field induced SOC, such as the Bychkov-Rashba SOC in two-dimensional heterolayers, may be feasible.
Determination of the Loschmidt echo envelope allows us to determine the "true", i.e. irreversible spin-relaxation time. The data shows an exponential time dependence of the Loschmidt echo envelope over several orders of magnitude (shown in Ref. \onlinecite{SupMat}) except for the beginning of the envelope for $\tau\ll 1$ where it starts with zero derivative due to geometric reasons.

\begin{figure}[htp]
\begin{center}
\includegraphics[scale=.5]{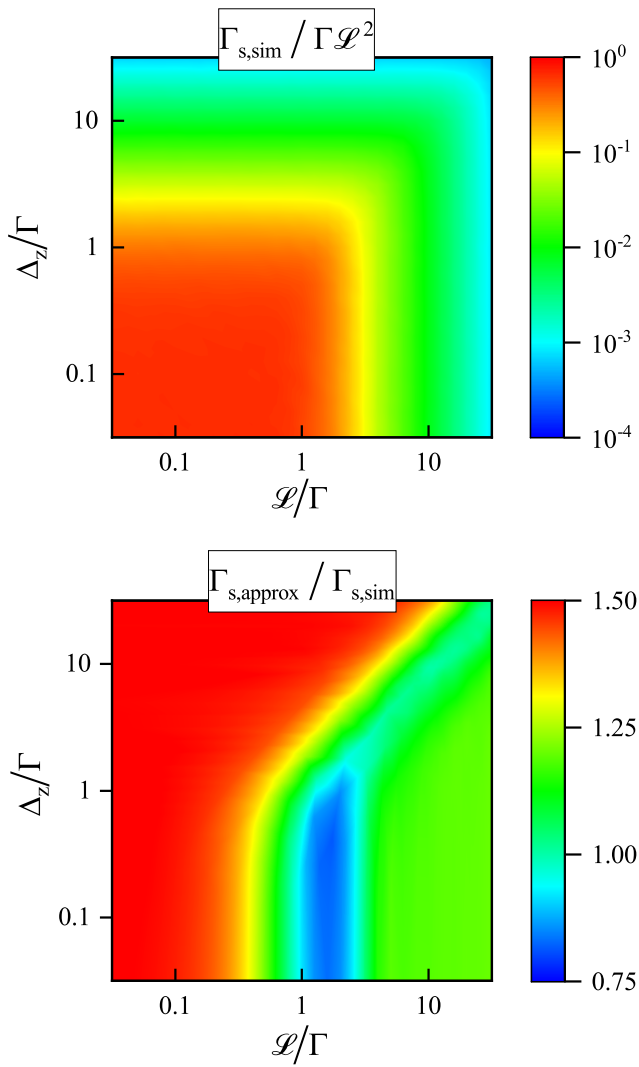}
\caption{\textit{Upper panel:} Phase diagram of the spin-relaxation rate, $\Gamma_{\text{s,sim}}$, for the Dresselhaus SOC as a function of $\mathcal{L}/\Gamma$ and $\Delta_{\text{Z}}/\Gamma$. The data are normalized by $\Gamma \mathcal{L}^2$. \textit{Lower panel:} Ratio of the approximate generic formula of $\Gamma_{\text{s,approx}}=\frac{\Gamma \mathcal{L}^2}{\Gamma^2+\mathcal{L}^2+\Delta_{\text{Z}}^2}$ and $\Gamma_{\text{s,sim}}$. Note that this ratio is around unity which justifies the use of the generic formula.}
\label{fig:square_plot_SR}
\end{center}
\end{figure}

\textit{The generic phase diagram of spin relaxation.} The calculated spin-relaxation rate, $\Gamma_{\text{s,sim}}$, is plotted for the entire phase diagram as a function of $\mathcal{L}/\Gamma$ and $\Delta_{\text{Z}}/\Gamma$ in the upper panel of Fig. \ref{fig:square_plot_SR}. $\Gamma_{\text{s,sim}}$ values are obtained by fitting exponential time dependences to the Loschmidt echo envelopes, which are obtained from our MC calculations. The data are normalized by $\Gamma \mathcal{L}^2$; without this normalization, the value of $\Gamma_{\text{s,sim}}$ changes by 8 orders of magnitude for the given range. We note that the cyclotron orbital motion of $\mathbf{k}$ in high magnetic field was neglected in the calculations, however we believe that this could be straightforwardly implemented.

The lower panel of Fig. \ref{fig:square_plot_SR}. shows the ratio of the suggested generic formula $\Gamma_{\text{s,approx}}=\frac{\Gamma \mathcal{L}^2}{\Gamma^2+\mathcal{L}^2+\Delta_{\text{Z}}^2}$ and $\Gamma_{\text{s,sim}}$ obtained from the MC calculations. A good agreement is found between the MC simulations and the generic formula for the entire phase diagram, the difference not being larger than a factor of 2. This strongly argues for the existence of a generic formula and for the validity of the presently suggested form. We believe that a similar formula is valid for an arbitrary SOC distribution while some multiplying factors (around unity) can be present in it.

We also verified the validity of the suggested generic formula for a specific spin-relation example which is exactly solvable. Burkov and Balents \cite{BurkovBalents} studied a two-dimensional electron gas with a lateral electric field, which induces a Bychkov-Rashba type SOC. They presented an \textit{analytic result} for the spin-relaxation rate using a many-body approach for arbitrary values of $\Gamma$, $\mathcal{L}$, and $\Delta_{\text{Z}}$. We found that our approximating formula well explains the analytic results for the entire phase diagram within a factor of 2. Details of the analytic calculations in the various regimes are somewhat involved and are presented in the Supplementary Material \cite{SupMat} along with the comparison figure between the analytic result and the approximation.   

This agreement provides an additional support for the validity of the recommended generic formula. We believe that besides the aforementioned multiplying factors, the formula may serve with a strong predicting value for the spin-relaxation and spin-transport \cite{Iordanskii_original,Iordanskii} properties in future spintronics materials. In addition, it describes well the general trends for the spin relaxation rate versus its parameters, which can help to identify the relevant model of a relaxation mechanism. Although we did not cover the case of spin relaxation in systems with inversional symmetry (the Elliott-Yafet theory \cite{Elliott,yafet1963g}), the recently discovered equivalence between the D'yakonov-Perel' and Elliott-Yafet Hamiltonians \cite{SzolnokiPRB2017} allows for a straightforward application of the present result for the Elliott-Yafet case.

%\section*{Conclusions}
\textit{Conclusions.} In conclusion, we studied the spin dynamics in zincblende semiconductors with the Dresselhaus spin-orbit coupling. We presented a model, which directly provides the quantum trajectories of individual spins and is equivalent to solving the spin Boltzmann equation. We identified a non-exponential, spin-dephasing regime of spin dynamics, which occurs due to a strong SOC. We tackled dephasing with the introduction of a Loschmidt echo. This allowed us to determine the spin-relaxation time for the entire spin-relaxation phase diagram involving the strength of quasiparticle scattering rate, spin-orbit coupling, and the Zeeman interaction. We found that a simple and compact form approximates well $\tau_{\text{s}}$. The validity of the formula was also confirmed for the two-dimensional electron gas with a lateral electric field (i.e. with a Bychkov-Rashba SOC) for which the spin-relaxation time could be determined analytically.

\section*{Acknowledgements}
The authors are indebted to Jaroslav Fabian for many stimulating discussions. Work supported by the Hungarian National Research, Development and Innovation Office (NKFIH) Grant Nrs. 2017-1.2.1-NKP-2017-00001, K124176, and K119442. A.K. acknowledges the Bolyai Program of the Hungarian Academy of Sciences.
%, and by the DFG SFB 1277
%\bibliographystyle{apsrev4}
%\bibliographystyle{apsrev}

%\bibliography{Tubes2011June_merged}

\appendix
\newpage
\pagebreak
\clearpage

\sloppy

\title{Supplementary Materials for: The generic phase diagram of spin relaxation in solids and the Loschmidt echo}

\author{G\'{a}bor Cs\H{o}sz}
\affiliation{Department of Physics, Budapest University of Technology and Economics and
MTA-BME Lend\"{u}let Spintronics Research Group (PROSPIN), POBox 91, H-1521 Budapest, Hungary}

\author{L\'enard Szolnoki}
\affiliation{Department of Physics, Budapest University of Technology and Economics and
MTA-BME Lend\"{u}let Spintronics Research Group (PROSPIN), POBox 91, H-1521 Budapest, Hungary}

\author{Annam\'aria Kiss}
\affiliation{Institute for Solid State Physics and Optics, Wigner Research Centre for Physics, Hungarian Academy of Sciences, POBox 49, H-1525 Budapest, Hungary}
\affiliation{Department of Physics, Budapest University of Technology and Economics and
MTA-BME Lend\"{u}let Spintronics Research Group (PROSPIN), POBox 91, H-1521 Budapest, Hungary}

\author{Bal\'azs D\'ora}
\affiliation{Department of Theoretical Physics, Budapest University of Technology and Economics and
MTA-BME Lend\"{u}let Topology Research Group (TOPCOR), POBox 91, H-1521 Budapest, Hungary}

\author{Jaroslav Fabian}
\affiliation{Institute for Theoretical Physics, University of Regensburg, 93040 Regensburg, Germany}

\author{Ferenc Simon}
\email{f.simon@eik.bme.hu} %[Corresponding author:]
\affiliation{Department of Physics, Budapest University of Technology and Economics and
MTA-BME Lend\"{u}let Spintronics Research Group (PROSPIN), POBox 91, H-1521 Budapest, Hungary}

\begin{widetext}

\section{Equivalence of spin Boltzmann equation and the Monte Carlo method}
In the main text, we motivated that the presented Monte Carlo method is equivalent to solving the spin Boltzmann equation. Herein, we present three additional examples to illustrate this equivalence numerically. 

\begin{figure}[htp]
    \begin{center}
      \includegraphics[width=0.5\linewidth]{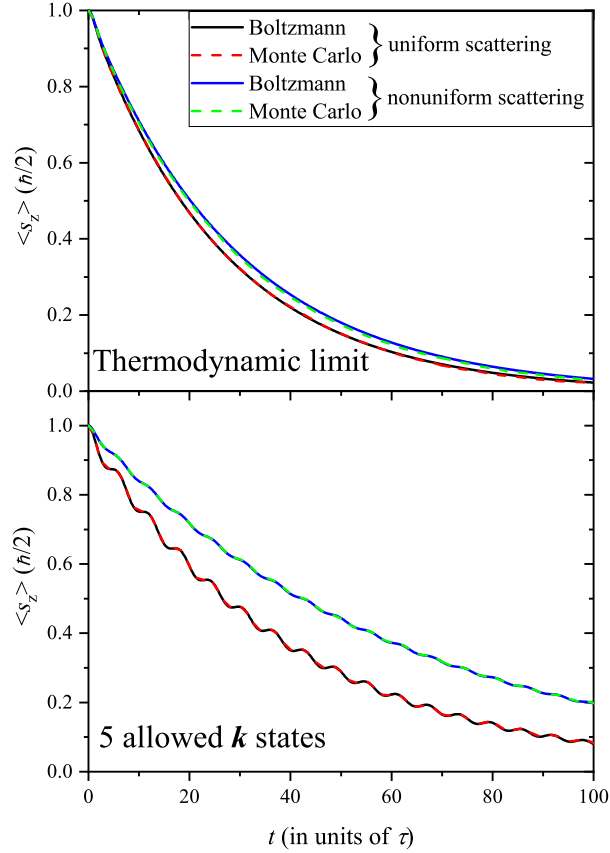}
      \caption{\textit{Upper panel:} Comparison of the numerical solution of spin Boltzmann equation and Monte Carlo based method for more than a thousand allowed $\vect{k}$ states, which is equivalent to the thermodynamic limit. $W_{\vect{k}\vect{k'}}$ is constant for all $\vect{k, k'}$ for the uniform scattering and it is weighted with $\text{cos}\theta$ ($\theta$ being the angle of $\vect{k}$ and the $z$ direction) in the nonuniform case. \textit{Lower panel:} The same comparison but the number of allowed $\vect{k}$ states is $5$, which are given in the text. For all calculations $\Delta=1$, $\Gamma=1$,\footnote{} and $\mathcal{L}=1$ was used.}
      \label{fig:SM_MC_Boltzmann_comparison}
    \end{center}
  \end{figure}

In Fig. \ref{fig:SM_MC_Boltzmann_comparison}., we compare the result of the Monte Carlo calculations with the solution of the spin Boltzmann equation. The upper panel gives two examples: one with a uniform scattering (i.e. $W_{\vect{k}\vect{k'}}$ constant for all $\vect{k, k'}$) and a nonuniform scattering. For the latter, $W_{\vect{k}\vect{k'}}$ is weighted by $\text{cos}\theta$, where $\theta$ is the angle between $\vect{k}$ and the $z$ direction. The calculations were performed for more than a thousand allowed $\vect{k}$ states which were spread uniformly over the Fermi surface. 

The lower panel of Fig. \ref{fig:SM_MC_Boltzmann_comparison}. shows the result of both types of calculations for 5 allowed $\vect{k}$ states only: the "North Pole" (NP), "South Pole" (SP), and 3 points along the "Equator" (E1, E2, and E3) which form an equilateral triangle on the Fermi surface. Again, $W_{\vect{k}\vect{k'}}$ is constant for the uniform scattering and $W_{\text{NP}\rightarrow \text{SP}}:W_{\text{NP/SP}\rightarrow \text{E}}=3:1$ for the nonuniform scattering modell.

Remarkably, we find no difference between solving the spin Boltzmann equation and the result of the Monte Carlo model for either case. Therefore the Monte Carlo model gives an accurate description of the spin dynamics, irrespective of the $\vect{k}$ distribution and the uniformity of the scattering. 

\section{The relation between the spin Boltzmann equation and the Bloch equations}

The phenomenological Bloch equations describe the time evolution of a spin ensemble under the action of an external DC magnetic field and an additional AC magnetic field. The latter is used in magnetic resonance experiments and is usually polarized perpendicularly to the DC field. The Bloch equations are usually written for the magnetization $\vect {M}$, which is the ensemble averaged magnitude of the spin magnetic dipole moments per unit volume. The Bloch equations read:

\begin{equation}
  \begin{aligned}
    \frac {\dd M_x(t)} {\dd t} &= \gamma ( \vect {M} (t) \times \vect {B} (t)  ) _x - \frac {M_x(t)} {T_2}, \\
    \frac {\dd M_y(t)} {\dd t} &= \gamma ( \vect {M} (t) \times \vect {B} (t)  ) _y - \frac {M_y(t)} {T_2}, \\
    \frac {\dd M_z(t)} {\dd t} &= \gamma ( \vect {M} (t) \times \vect {B} (t)  ) _z - \frac {M_z(t) - M_0} {T_1},
  \end{aligned}
\end{equation}

Here, the Larmor vector due to the external magnetic fields (either DC or AC or both) is identified as $\vect{\Omega(t)}=\gamma \vect {B} (t)$, where $\gamma$ is the so-called gyromagnetic ratio for electrons. The $T_1$ and $T_2$ are the so-called longitudinal and transversal relaxation times. $M_0$ appears when the DC magnetic field is along the $z$ axis. The phenomenological relaxation times describe that following an excitation, the respective magnetizations return to their equilibrium values, which is $M_0$ for the $M_z$ component and 0 for $M_x$ and $M_y$. In zero external external magnetic field, the $T_1$ and $T_2$ distinction vanishes. 

We recognize a clear analogy between the spin Boltzmann equation and the Bloch equations. 
Although the spin Boltzmann equation contains the external magnetic field and the $\mathbf{k}$ dependent built-in SOC related magnetic fields altogether, the latter give rise to the phenomenological relaxation times ($T_1$ and $T_2$) in combination with the momentum relaxation events (which are described by the $W_{\mathbf{k'}\mathbf{k}}$ terms in the spin Boltzmann equation. The external magnetic field appears unchanged in the Bloch equations. 

Strictly speaking, this is only valid for the case when the spin magnetization decays exponentially according to the spin Boltzmann equation. However, the effect of dephasing can be also included in the Bloch equations by introducing e.g. spatial dependence of the local DC magnetic fields (e.g. due to the inhomogeneity of the magnet) or particle orientation dependent $\gamma$ (or $g-$factor) in a powder sample. The spin-Boltzmann equation can also be amended with the diffusion term, whose analogue is known as the Bloch-Torrey equations. 

\section{The Loschmidt echo in magnetic resonance}

We demonstrate herein the generic concept of the Loschmidt echo for the case of magnetic resonance. Therein, the so-called spin echo is a specific case of a Loschmidt echo.

The dephasing problem is encountered in magnetic resonance and it is also tackled with a version of the Loschmidt echo. It is a common challenge in magnetic resonance that dephasing and spin relaxation processes are simultaneously present. The so-called spin-echo is employed to tackle this problem. 

Most generally, one encounters three different time scales in magnetic resonance: $T_2^*$, $T_2$, and $T_1$. Of these, $T_{1,2}$ are irreversible relaxation processes and $T_2^*$ is related to the reversible dephasing processes \cite{AbragamBook,SlichterBook}. 
The distinction between $T_1$ and $T_2$ stems from the fact that a magnetic field is applied, which inevitably leads to a distinction between relaxation processes for the magnetization components which are parallel (the $T_1$ processes, also known as longitudinal relaxation time) and perpendicular ((the $T_2$ processes, also known as transversal relaxation time)) to the external magnetic field. In zero magnetic field, this distinction vanishes.

  \begin{figure}[htp]
    \begin{center}
      \includegraphics[width=0.5\linewidth]{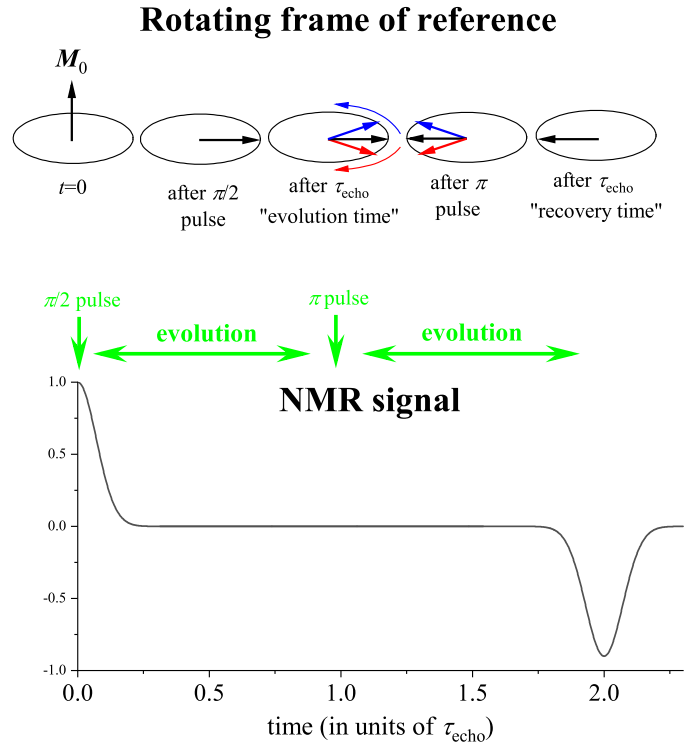}
      \caption{Schematics of the dephasing process in NMR experiments and the method of spin-echo. The figure assumes a right handed precession direction with $\overline{\omega_{\text{L}}}$. The spin magnetization lies in the $(x',y')$ plane after a $\pi/2$ pulse when dephasing due to a spread in the Larmor frequencies starts: spins in blue and red precess faster or slower than $\overline{\omega_{\text{L}}}$, respectively. After an evolution time of $\tau_{\text{echo}}$, a $\pi$ pulse is applied which rotates the spins around an axis perpendicular to $z'$. Clearly, the blue and red spins are now behind or before the average spin direction and as a result these will be aligned coherently after another $\tau_{\text{echo}}$ time, when the spin echo occurs. The lower panel depicts the corresponding NMR signal.}
      \label{fig:NMR_schematics}
    \end{center}
  \end{figure}

The Bloch equations \cite{SlichterBook} describe the motion of spins in a DC magnetic field along the $z$ axis, which is accompanied by an AC magnetic field whose polarization rotates around $z$. In equilibrium, the magnetization of the spin ensemble, $\bf M$ is stationary along the $z$ axis with a value of $\bf M_0$. When the AC magnetic field is applied in a pulsed manner, the magnetization is rotated away from the $z$ axis and starts to precess around $z$ with the Larmor frequency $\omega_{\text{L}}=\gamma B$ ($B$ is the magnetic field and $\gamma$ is the so-called gyromagnetic ratio of the studied spin system, e.g. $\gamma \approx 2\pi 42.6\text{MHz/T}$ for protons and $\gamma \approx 2\pi 28.0\text{GHz/T}$ for electrons). 

In a typical experiment, an AC irradiation is applied with an angular-frequency matching $\omega_{\text{L}}$ and a pulse duration which is sufficient to rotate $\bf M$ into the $(x,y)$ plane. This is known as a $\pi/2$ pulse, as the magnetization is rotated perpendicular to $z$. Then the $(x,y)$ and $z$ components of the non-equilibrium spin magnetization decay to the respective equilibrium values (0 and $M_0$)with $T_2$ and $T_1$ relaxation times. However, in most cases the $(x,y)$ components vanish much earlier than $T_2$ due to dephasing: local magnetic field inhomogeneities are present which lead to a distribution in $\omega_{\text{L}}$. The inhomogeneities are caused by either defects or impurities (these are the leading cause in solid state NMR) or by the inevitable inhomogeneity of the magnet (this is the leading cause in high resolution or liquid NMR) \cite{AbragamBook}.

  \begin{figure}[htp]
    \begin{center}
      \includegraphics[width=.5\linewidth]{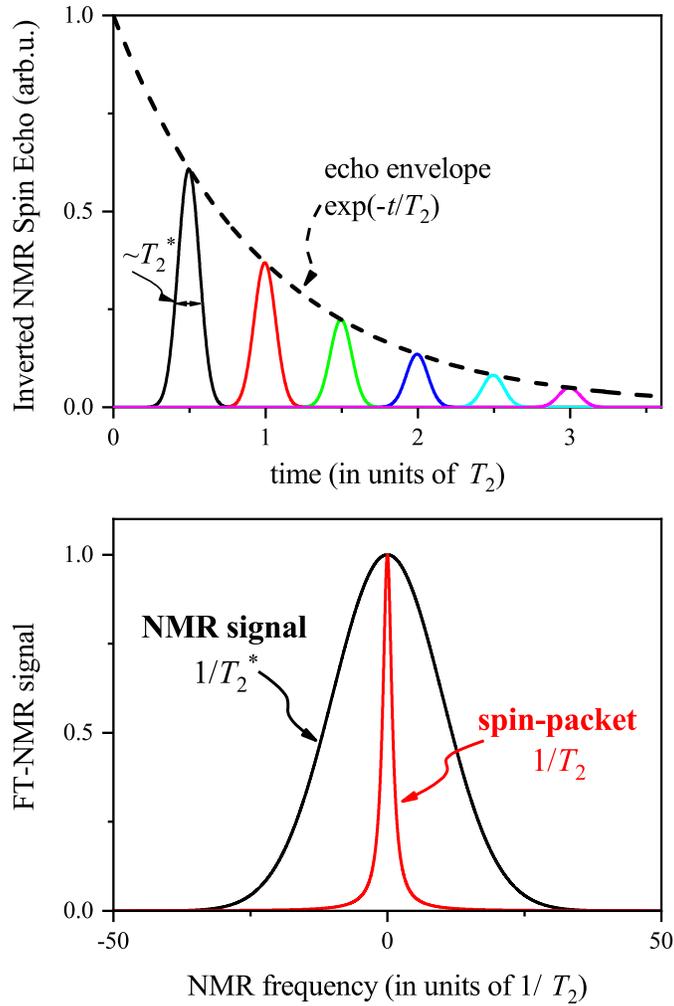}
      \caption{Schematics of the $T_2$ measurement. NMR spin echo experiments are performed with varying time delay between the $\pi/2$ and $\pi$ pulses. The individual echoes have a linewidth of $2T_2^*$ but the resulting spin echo envelope follows $\text{e}^{-t/T_2}$. The corresponding FT NMR signal reflects this behavior: it contains a broad signal whose width is $~1/T_2^*$ and it consists of individual spin-packets whose width is $~1/T_2$.}
      \label{fig:NMR_envelope_schematics}
    \end{center}
  \end{figure}

This process is usually described in a frame of reference which rotates with the mean value of the Larmor (angular frequency), $\overline{\omega_{\text{L}}}$ around the $z$ axis and is schematically shown in Fig. \ref{fig:NMR_schematics}. The coordinate axes of the rotating frame of reference are denoted by $x'$, $y'$, and $z'$ ($z'$ is identical to the $z$ axis). The originally $M_0$ magnetization lies in the $(x',y')$ plane after a $\pi/2$ pulse where dephasing starts. In the rotating frame of reference, some spins have angular frequencies which are larger (the blue arrows in the figure) or smaller (the red arrows in the figure) than $\overline{\omega_{\text{L}}}$. The resulting net magnetization vanishes on a timescale of $T_2^*\approx 1/\Delta \omega_{L}$, where the latter is the spread in the Larmor frequencies. Another pulse is applied after a so-called "evolution time", $\tau_{\text{echo}}$, which rotates the spins by $\pi$. The figure depicts the location of the spins which precess faster (blue arrows) and slower (red arrows) than the average after the $\pi$ pulse. Clearly, after a waiting time of another $\tau_{\text{echo}}$ the spins are aligned again coherently in the $x',y'$ plane with a direction opposite to their original coherent direction. 

Fig.~\ref{fig:NMR_schematics}. also shows the corresponding NMR signal: the initially decaying signal is partially recovered, i.e. an echo is observed at $2\tau_{\text{echo}}$ when the $\pi$ pulse is applied at $\tau_{\text{echo}}$. The reason why the NMR spin echo is observed, is that the dephasing is not accompanied by a memory loss, thus each spin "remembers" the magnitude of its Larmor frequency. However in reality, memory loss is also present on the spin-relaxation timescale, $T_2$, where typically $T_2^*\gg T_2$. In NMR, the physical origin of $T_2$ can be dipole-dipole interaction (this is the leading mechanism in solid state NMR) or molecular diffusion (this is the leading mechanism in high resolution or liquid NMR) \cite{AbragamBook}.

Fig.~\ref{fig:NMR_envelope_schematics}. shows the schematics of the $T_2$ measurement: spin-echo experiments are performed consecutively (in different pulse sequence runs, each starting from the equilibrium $M_0||z'$) with varying $\tau_{\text{echo}}$. The envelope of the observed echoes follow $\text{exp}^{-\tau_{\text{echo}}/T_2}$, which allow for the determination of $T_2$, which is a true, irreversible relaxation time, clearly distinguishable from dephasing. After Fourier transformation, the NMR signal has a large linewidth of $~1/T_2^*$ (in frequency units) which consists of so-called spin-packets\cite{PortisPR1953}, whose linewidth is $~1/T_2$.

\section{Efficient calculation of the Loschmidt echo envelope in our numerical studies}

We outlined in the main text how individual Loschmidt echoes can be obtained by inverting the SOC related Larmor precession vectors. In principle, the envelope could be obtained from such individual echoes by varying the time delay of the reversal. Clearly, this procedure requires to calculate a full ensemble averaged time evolution and repeating this calculation over and over for each time delay points. However, it turns out that the envelope itself can be obtained more effectively when we are not interested in the individual Loschmidt echoes. It turns out that this calculation is not more time consuming than calculating a single time decay of the spin ensemble. 

 \begin{figure}[htp]
    \begin{center}
      \includegraphics[width=.5\linewidth]{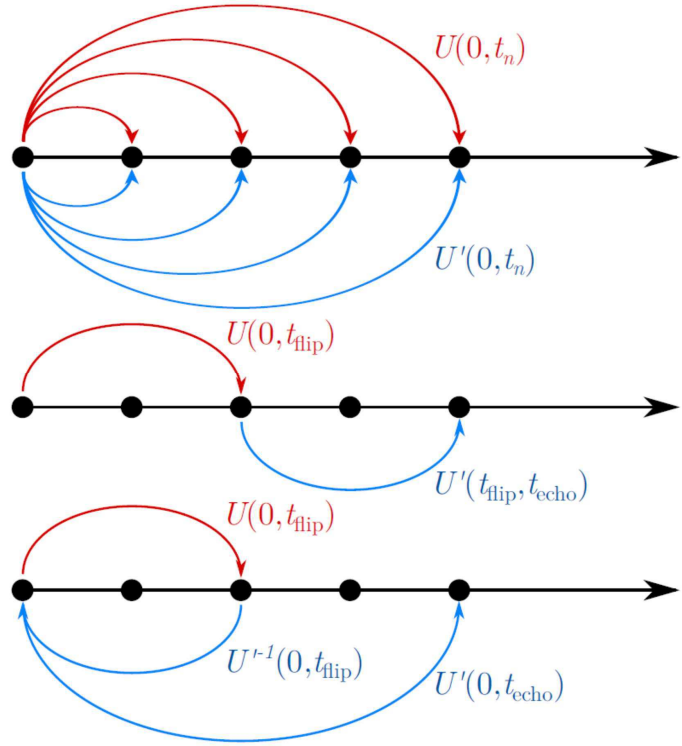}
      \caption{Schematics of the efficient Loschmidt echo envelope calculation. Top panel: the Larmor precession acts as a transformation operator, $U(t_1,t_2)$ on an individual spin. This can be obtained for an equidistant array of time values, even in the presence of momentum scattering which provides a new random $\mathbf{k}$ that alters the spin precession direction. $U'(t_1,t_2)$ is obtained for the \emph{inverted} Larmor precession vectors. Middle panel: the time trace of an individual echo with flip time $t_{\text{flip}}$ could be obtained at any time $t$ by acting on the spin with $U(0,t_{\text{flip}})$, followed by $U'(t_{\text{flip}},t)$. Bottom panel: the echo envelope at a time point $t_{\text{echo}}=2t_{\text{flip}}$ is obtained from the product $U(0,t_{\text{flip}})\times U'^{-1}(0,t_{\text{flip}}\times U'(0,t_{\text{echo}}))$.}
      \label{fig:Loschmidt_calculation_schematics}
    \end{center}
  \end{figure}

The schematics of the method is depicted in Fig. \ref{fig:Loschmidt_calculation_schematics}. It is based on keeping track of the rotation operator (which is a $2\times 2$ matrix), $U(t_1,t_2)$ which describes the evolution of a single spin at $t_1$ to a time point of $t_2$. Although momentum scattering happens in random time intervals, $U(t_1,t_2)$ can be constructed for any $t_1$ and $t_2$, which also involves the randomizing nature of the momentum scattering, which changes the direction of the Larmor precession. However, it is practical to predefine an equidistant array of time steps for which the envelope is to be calculated.

In addition to keeping track of the rotation operator under the action of the Larmor precession with randomized $\mathbf{k}$ values, we can keep track of the rotation operators which would act for the \emph{inverted} Larmor precession vectors. This is denoted by $U'(t_1,t_2)$. The top panel of Fig. \ref{fig:Loschmidt_calculation_schematics} depicts by arrows the action of these two types of operators. 

Next, we consider an individual Loschmidt echo where the SOC-related Larmor vectors are inverted at a flip time of $t_{\text{flip}}$. The middle panel in Fig. \ref{fig:Loschmidt_calculation_schematics} depicts that the echo can be obtained for any arbitrary time $t$from the action of $U(0,t_{\text{flip}})$, followed by $U'(t_{\text{flip}},t)$, i.e. their product.
The bottom panel describes the efficient method to obtain the Loschmidt echo envelope at an arbitrary $t_{\text{echo}}=2t_{\text{flip}}$. It requires the subsequent action of $U(0,t_{\text{flip}})$, $U'^{-1}(0,t_{\text{flip}}$ and $U'(0,t_{\text{echo}}))$ since the identity:

\begin{gather}
U(0,t_{\text{flip}})\times U'(t_{\text{flip}},2 t_{\text{flip}})=U(0,t_{\text{flip}})\times U'^{-1}(0,t_{\text{flip}} \times U'(0,2 t_{\text{flip}}))
\end{gather}
holds. Clearly this method involves a larger memory use but it substantially accelerates the calculations. In the end, the Loschmidt echo envelope for each individual spins needs to be ensemble averaged to obtain the final result.

%\section{Long time dependence of the Loschmidt echo envelope}

  \begin{figure}[htp]
    \begin{center}
      \includegraphics[width=.5\linewidth]{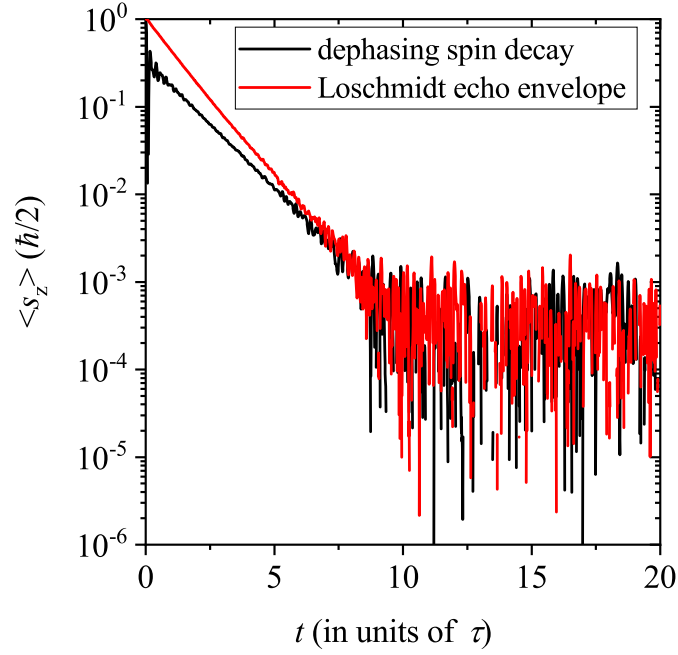}
      \caption{Time dependence of the Loschmidt echo envelope and the spin decay signal for longer times on a semilog plot. Note the exponential decay for both the spin-decay and the Loschmidt echo envelope signals.}
      \label{fig:SM_Loschmidt_Longtime}
    \end{center}
  \end{figure}

Fig. \ref{fig:SM_Loschmidt_Longtime}. shows the time dependence of the Loschmidt echo envelope and the spin decay signal for longer times on a semilog plot (vertical axis is logarithmic). Note that both signals decay exponentially for longer times. The apparent noise in the signals for longer times could in principle be reduced by increasing the ensemble.

\section{Spin relaxation for a two-dimensional electron gas with the Bychkov-Rashba spin-orbit coupling}

Burkov and Balents investigated the spin relaxation of a 2DEG with Rashba SOC.

\begin{equation}
  H_0 =
    \frac{\vect{\pi}^2}{2m}
    + \lambda \hat{z}\cdot \left[ \vect{\sigma} \times \vect{\pi} \right]
    - \frac{\Delta_\mathrm{Z}}2 \sigma^z,
  \label{SM_eq:BB_ham_orig}
\end{equation}

\noindent where $\vect{\pi} = \vect{p} + (e/c)\vect{A}$ is the kinetic momentum,
  $\vect{\sigma}$ is the spin operator (Pauli-matrices), and $\Delta_\mathrm{Z} = g \mu_\mathrm{B} B$ is the Zeeman energy.

The following substitutions are used to convert to our notation.

\begin{equation}
  \begin{aligned}
    \mathcal{L} &= 2\lambda p_\mathrm{F}, \\
    \vect{s}    &= \vect{\sigma}/2,
  \end{aligned}
\end{equation}

\noindent where $p_\mathrm{F}$ is the absolute value of the momentum at the Fermi-energy.

The Hamiltonian can be written after substitution:

\begin{equation}
  H_0 =
    \frac{\vect{\pi}^2}{2m}
    + \mathcal{L} \hat{z}\cdot \left[ \vect{s} \times \frac{\vect{\pi}}{p_\mathrm{F}} \right]
    - \Delta_\mathrm{Z} s^z.
  \label{SM_eq:BB_ham}
\end{equation}

Burkov and Balents defined the following intermediate quantities to simplify their equations:

\begin{equation}
  \begin{aligned}
    \Delta &=
      2 \sqrt{
        \left(
          \frac{
            \hbar \omega_c
            + \Delta_\mathrm{Z}
          }2
        \right)^2
        + 2 \lambda^2 m \varepsilon_\mathrm{F}
      }, \\
    \cos\vartheta &=
      \sqrt{
        \frac12
        - \frac{
            \hbar\omega_c
            + \Delta_\mathrm{Z}
          }{
            2 \Delta
          }
      }.
  \end{aligned}
\end{equation}
Here $\omega_c$ denotes the cyclotron angular frequency.

Using self-consistent Bohr approximation (SCBA), they calculated the spin-diffusion propagator.

\begin{equation}
  \begin{aligned}
    D_{zz}^{-1}(\Omega) =&
     1
     - f_0(\Omega)
     + \frac{
         \sin^2(2\vartheta)
       }2
       \left[
         2f_0(\Omega)
         - f_+(\Omega)
         - f_-(\Omega)
       \right], \\
    D_{+-}^{-1}(\Omega) =&
      1
      - \frac{
          \sin^2(2\vartheta)
        }2
        f_0(\Omega+\omega_c)
      - \sin^4(\vartheta)
        f_+(\Omega+\omega_c) \\
     &- \cos^4(\vartheta)
        f_-(\Omega+\omega_c),
  \end{aligned}
  \label{eq:BB_diff_propagator}
\end{equation}

\noindent where

\begin{equation}
  \begin{aligned}
    f_0(\Omega) &=
      \frac1{
        1-i\Omega\tau
      }, \\
    f_{\pm}(\Omega) &=
      \frac1{
        1-i\Omega\tau\pm i\Delta\tau
      }.
  \end{aligned}
\end{equation}

The spin-relaxation times can be calculated using the poles of the spin-diffusion propagators,
  or, equivalently, zeroth of the inverse diffusion propagators as written above.

These equations can be further simplified by introducing the quantity $\Delta_m = \hbar\omega_c + \Delta_\mathrm{Z}$.

\begin{equation}
  \begin{aligned}
    \Delta &=
      2 \sqrt{
          \left(
            \frac{
              \hbar \omega_c
              + \Delta_\mathrm{Z}
            }2
          \right)^2
          + 2 \lambda^2 m \varepsilon_\mathrm{F}
        } \\
      &=
      2 \sqrt{
          \frac{
            \Delta_m^2
          }4
          + \cancel{2} \lambda^2 \cancel{m} \frac{p_\mathrm{F}^2}{\cancel{2m}}
        } \\
      &=
        \sqrt{
          \Delta_m^2 + \mathcal{L}^2
        }, \\
    \cos\vartheta &=
      \sqrt{
        \frac12
        - \frac{
            \hbar\omega_c
            + \Delta_\mathrm{Z}
          }{
            2 \Delta
          }
      } \\
      &=
      \sqrt{
        \frac12
        - \frac{\Delta_m}{2\Delta}
      }, \\
    \sin\vartheta &= 
      \pm \sqrt{1-\cos^2\vartheta} \\
      &=
      \pm \sqrt{
            \frac12
            + \frac{\Delta_m}{2\Delta}
          }, \\
    \sin^2(2\vartheta) &=
      \left(2 \sin\vartheta \cos\vartheta\right)^2 \\
      &= \frac{\mathcal{L}^2}{\Delta^2}.
  \end{aligned}
\end{equation}

Substituting these results into the spin-diffusion propagator:

\begin{equation}
  D_{zz}^{-1}(\Omega) =
    \frac{
      \mathcal{L}^2\tau^2
      - i \tau \left( 1 + \Delta^2 \tau^2 \right) \Omega
      - 2 \tau^2 \Omega^2
      + i \tau^3 \Omega^3
    }{
      \left(
        1 - i \Omega \tau
      \right)
      \left(
        1 - i \Omega \tau + i \Delta \tau
      \right)
      \left(
        1 - i \Omega \tau - i \Delta \tau
      \right)
    }
\end{equation}

After substituting $\Gamma = 1/\tau$ and $\Omega = - i s$ (Laplace domain): 

\begin{equation}
  D_{zz}^{-1}(s) =
    \frac{
      - \Gamma  \mathcal{L}^2
      + \left(\Gamma ^2+\Delta_m^2+\mathcal{L}^2\right) s
      - 2 \Gamma s^2
      + s^3
    }{
      (s-\Gamma )
      \left(s-\Gamma -i \sqrt{\Delta_m^2+\mathcal{L}^2}\right)
      \left(s-\Gamma +i \sqrt{\Delta_m^2+\mathcal{L}^2}\right)
    }.
  \label{eq:Diff_prop}
\end{equation}

The numerator of $D_{zz}^{-1}(s)$ is a third degree, real coefficient polynomial for $s$.
The roots of this polynomial are the poles of the diffusion propagator.
The real parts of the roots are the inverses of the respective relaxation times of the corresponding exponential relaxation. 

\begin{equation}
  \tau_{\mathrm{s},i} = 1/\Re[s_i].
\end{equation}

Although the roots of a cubic polynomial can be found using Cardano's method,
  the resulting expressions for the roots are hard to interpret.

\begin{equation}
  \begin{aligned}
    \Delta_0 &= \Gamma^2 - 3\left( \mathcal{L}^2 + \Delta_m^2 \right), \\
    \Delta_1 &= 9 \Gamma \mathcal{L}^2 - 2 \Gamma^3 - 18 \Gamma \Delta_m^2, \\
           C &= \left(
                  \frac{
                    \Delta_1
                    + \sqrt{ \Delta_1^2 - 4 \Delta_0^3 }
                  }{
                    2
                  }
                \right)^{\frac13}, \\
    \varepsilon_3 &= -\frac12 + \frac{\sqrt{3}}2 i, \\
    s_i &= -\frac13
              \left(
                2 \Gamma
                + \varepsilon_3^k C
                + \frac{
                    \Delta_0
                  }{
                    \varepsilon_3^k C
                  }
              \right)
        ,\quad k \in \{0,1,2\}.
  \end{aligned}
\end{equation}

The most relevant relaxation time out of the three is the longest,
  which corresponds to the pole with the smallest real part.

\begin{equation}
  \Gamma_s = \min\{ \Re[s_i] \}.
\end{equation}

Taking the poles in specific limits yields more interpretable results.

Perturbation theory for transcendental equations can be used for calculating the roots at different regimes.
Although the equation is not transcendental,
  it's more effective to get the roots of a quadratic polynomial using perturbation theory in different regimes than applying the limits to the exact expressions for the roots.

\subsection{A perturbative treatment of the transcendental equations}

An equation in the following form is assumed:

\begin{equation}
  g(x,\lambda) = 0.
\end{equation}

Denote the roots for this equations as $x_i$.
The equation is assumed to be easy to solve for $\lambda = 0$,
  but it becomes hard or impossible to solve at other $\lambda$ values.
The goal is getting $x_i$ as a series expansion of $\lambda$.
In the following $x_i$ are treated as functions of $\lambda$.

\begin{equation}
  \begin{aligned}
    g(x_i(\lambda), \lambda) &= 0, \\
    \frac{\partial g}{\partial x}\Bigr|_{x_i, \lambda} \frac{\d x_i}{\d \lambda}
    + \frac{\partial g}{\partial \lambda}\Bigr|_{x_i, \lambda}
    &= 0, \\
    \frac{\d x_i}{\d \lambda}
    &= -\frac{\partial g}{\partial \lambda}\Bigr|_{x_i, \lambda}
        \left(
          \frac{\partial g}{\partial x}\Bigr|_{x_i, \lambda}
        \right)^{-1}.
  \end{aligned}
\end{equation}

Evaluate at $\lambda = 0$.

\begin{equation}
  \begin{aligned}
    g(x_i(0), 0) &= 0, \\
    \frac{\d x_i}{\d \lambda}\Bigr|_{\lambda=0}
    &= -\frac{\partial g}{\partial \lambda}\Bigr|_{x_i(0), 0}
        \left(
          \frac{\partial g}{\partial x}\Bigr|_{x_i(0), 0}
        \right)^{-1}, \\
    x_i(\lambda)
    &= x_i(0)
       + \frac{\d x_i}{\d \lambda}\Bigr|_{\lambda=0} \lambda
       + \mathcal{O}( \lambda^2 ).
  \end{aligned}
\end{equation}

Higher order approximations can be get by applying higher order derivatives to the equation.
The second derivative of $x_i(\lambda)$:

\begin{equation}
  \frac{\d^2 x}{\d \lambda^2}
  = - \left(
        \frac{\partial g}{\partial x}
      \right)^{-1}
      \left[
        \frac{\partial^2 g}{\partial x^2}
          \left(
            \frac{\d x_i} {\d \lambda}
          \right)^2
        + 2 \frac{\partial^2 g}{\partial x \partial \lambda}
          \frac{\d x_i} {\d \lambda}
        + \frac{\partial^2 g}{\partial \lambda^2}
      \right],
\end{equation}

In the following sections only first order perturbation is calculated,
  while the second derivative is used to estimate the Lagrange remainder of the Taylor expansion.

\subsection{The $\mathcal{L} \ll \max(\Gamma,\Delta_m)$ regime}

The poles of the spin diffusion operator are the roots of the numerator of Eq.~\ref{eq:Diff_prop}.
First order perturbation is applied substituting $x=s$ and $\lambda=\mathcal{L}^2$.
First the roots at $\lambda=0$ have to be found.

\begin{equation}
  \begin{aligned}
    -s_i(0)\left(\Gamma^2+\Delta_m^2\right)
    + 2 s_i(0)^2 \Gamma
    - s_i(0)^3
    &= 0, \\
    s_1(0) &= 0, \\
    s_{2,3}(0) &= \Gamma \pm i \Delta_m.
  \end{aligned}
\end{equation}

At this regime only the first root is relevant as its' real part is much smaller than the other two roots.
The first order perturbation result for this root:

\begin{equation}
  s_1 = \frac{ \mathcal{L}^2 }{\Gamma^2 + \Delta_m^2}\Gamma
        + \mathcal{O} \left(
            \left(
              \frac{ \mathcal{L}^2 }{\Gamma^2 + \Delta_m^2}
            \right)^2
            \Gamma
          \right)
\end{equation}

This expression for the root is valid in the regime where the ratio of the error and the main term is much smaller than $1$.

\begin{equation}
  \begin{aligned}
    \frac{ \mathcal{L}^2 }{\Gamma^2 + \Delta_m^2} &\ll 1, \\
    \mathcal{L}^2 &\ll \Gamma^2 + \Delta_m^2, \\
    \mathcal{L}^2 &\ll 2\max(\Gamma^2,\Delta_m^2), \\
    \mathcal{L} &\ll \max(\Gamma,\Delta_m).
  \end{aligned}
\end{equation}

\subsection{The ``Small'' $\Gamma$ regime}

First order perturbation is applied again by substituting $x=s$ and $\lambda=\Gamma$.
The roots at $\lambda=0$.

\begin{equation}
  \begin{aligned}
    s_i(0) \left(
        -\Delta_m^2
        -\mathcal{L}^2
      \right)
    - s_i(0)^3
    &= 0, \\
    s_1(0) &= 0, \\
    s_{2,3}(0) &= \pm i \sqrt{\mathcal{L}^2 + \Delta_m^2}.
  \end{aligned}
\end{equation}

In this case all roots are relevant as all of them have the comparable ($0$) real part.

The first order correction of the first root:

\begin{equation}
  s_1
  = \frac{\mathcal{L}^2}{\mathcal{L}^2 + \Delta_m^2} \Gamma
    + \mathcal{O} \left(
        \frac{
          \mathcal{L}^4 \Gamma^3
        }{
          \left(
            \mathcal{L}^2
            + \Delta_m^2
          \right)^3
        }
      \right).
\end{equation}

Again, the regime of validity can be determined by examining the ratio of the main and error terms.
In this case the first order perturbation is valid when $\Gamma \ll \mathcal{L}$ or when $\Gamma\mathcal{L} \ll \Delta_m^2$.

For the oscillating roots it's only necessary to calculate the perturbation for one of them as they are conjugate pairs.
First order correction is only present for the real part.

\begin{equation}
  \begin{aligned}
    \Re[s_2]
    &= \frac{\mathcal{L}^2 + 2\Delta_m^2}{2(\mathcal{L}^2+\Delta_m^2)}\Gamma
      + \mathcal{O} \left(
        \frac{
          \Gamma^3
        }{
          \mathcal{L}^2
          + \Delta_m^2
        }
      \right), \\
    \Im[s_2]
    &= i \sqrt{\mathcal{L}^2 + \Delta_m^2}
       + \mathcal{O} \left(
           \frac{
             \Gamma^2
           }{
             \sqrt{\mathcal{L}^2 + \Delta_m^2}
           }
         \right).
  \end{aligned}
\end{equation}

These expressions are valid in the regime $\Gamma \ll \max(\mathcal{L},\Delta_m)$.

\subsection{The two regimes for the root with zero real part}

The expression for the first root for the two regimes can be combined in a single expression.

\begin{equation}
  s_1 = \frac{\mathcal{L}^2 \Gamma}{\mathcal{L}^2 + \Gamma^2 + \Delta_m^2}
        \left(
          1
          + \mathcal{O} \left(
              \frac{ \min(\mathcal{L},\Gamma)^2 }{
                \mathcal{L}^2
                + \Gamma^2
                + \Delta_m^2
              }
            \right)
        \right)
\end{equation}

The approximation is valid in all regimes where the error term is much smaller than $1$.

The error term is not negligible when $\mathcal{L} \simeq \Gamma$ and $\Delta_m \lesssim \mathcal{L}, \Gamma$.

Note that even if this expression is valid on a significant regime for the non-oscillating root,
there is a large regime where the oscillating roots have longer relaxation time compared to this root.

\begin{figure}[htp]
\begin{center}
\includegraphics[scale=0.5]{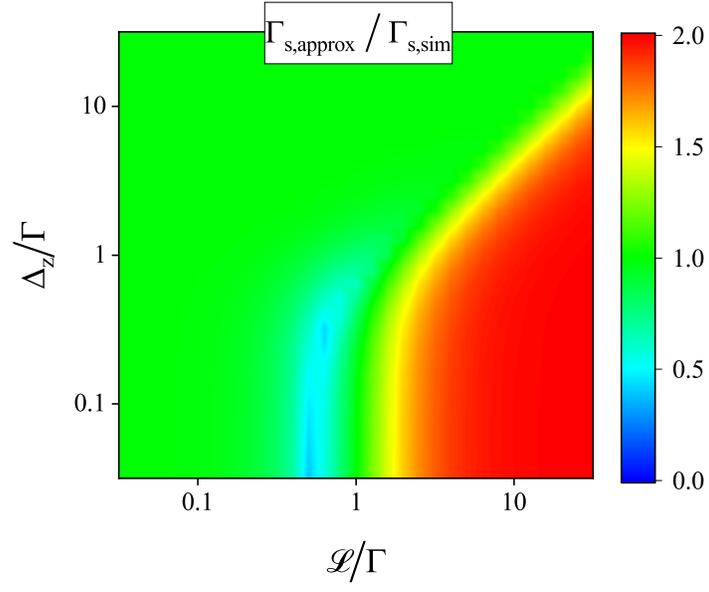}
\caption{Ratio of the recommended generic formula for the spin-relaxation rate and that obtained analytically for the 2D electron gas with the Bychkov-Rashba SOC.}
\label{fig:SM_TernaryPlot_2D_BychkovRashba}
\end{center}
\end{figure}

We finally show the ratio between the approximation formula in the main text and the herein presented simulated values in Fig. \ref{fig:SM_TernaryPlot_2D_BychkovRashba}. The agreement between the two kinds of data is close to unity and not deviating from it more than 50~\%.

%\bibliographystyle{naturemag}
%\bibliography{Tubes2011June_new}

\begin{thebibliography}{35}
\expandafter\ifx\csname natexlab\endcsname\relax\def\natexlab#1{#1}\fi
\expandafter\ifx\csname bibnamefont\endcsname\relax
  \def\bibnamefont#1{#1}\fi
\expandafter\ifx\csname bibfnamefont\endcsname\relax
  \def\bibfnamefont#1{#1}\fi
\expandafter\ifx\csname citenamefont\endcsname\relax
  \def\citenamefont#1{#1}\fi
\expandafter\ifx\csname url\endcsname\relax
  \def\url#1{\texttt{#1}}\fi
\expandafter\ifx\csname urlprefix\endcsname\relax\def\urlprefix{URL }\fi
\providecommand{\bibinfo}[2]{#2}
\providecommand{\eprint}[2][]{\url{#2}}

\bibitem[{\citenamefont{Wolf et~al.}(2001)\citenamefont{Wolf, Awschalom,
  Buhrman, Daughton, von Moln{\'a}r, Roukes, Chtchelkanova, and
  Treger}}]{WolfSCI}
\bibinfo{author}{\bibfnamefont{S.~A.} \bibnamefont{Wolf}},
  \bibinfo{author}{\bibfnamefont{D.~D.} \bibnamefont{Awschalom}},
  \bibinfo{author}{\bibfnamefont{R.~A.} \bibnamefont{Buhrman}},
  \bibinfo{author}{\bibfnamefont{J.~M.} \bibnamefont{Daughton}},
  \bibinfo{author}{\bibfnamefont{S.}~\bibnamefont{von Moln{\'a}r}},
  \bibinfo{author}{\bibfnamefont{M.~L.} \bibnamefont{Roukes}},
  \bibinfo{author}{\bibfnamefont{A.~Y.} \bibnamefont{Chtchelkanova}},
  \bibnamefont{and} \bibinfo{author}{\bibfnamefont{D.~M.}
  \bibnamefont{Treger}}, \bibinfo{journal}{Science}
  \textbf{\bibinfo{volume}{294}}, \bibinfo{pages}{1488} (\bibinfo{year}{2001}).

\bibitem[{\citenamefont{\ifmmode \check{Z}\else
  \v{Z}\fi{}uti\ifmmode~\acute{c}\else \'{c}\fi{}
  et~al.}(2004)\citenamefont{\ifmmode \check{Z}\else
  \v{Z}\fi{}uti\ifmmode~\acute{c}\else \'{c}\fi{}, Fabian, and
  Das~Sarma}}]{FabianRMP}
\bibinfo{author}{\bibfnamefont{I.}~\bibnamefont{\ifmmode \check{Z}\else
  \v{Z}\fi{}uti\ifmmode~\acute{c}\else \'{c}\fi{}}},
  \bibinfo{author}{\bibfnamefont{J.}~\bibnamefont{Fabian}}, \bibnamefont{and}
  \bibinfo{author}{\bibfnamefont{S.}~\bibnamefont{Das~Sarma}},
  \bibinfo{journal}{Rev. Mod. Phys.} \textbf{\bibinfo{volume}{76}},
  \bibinfo{pages}{323} (\bibinfo{year}{2004}).

\bibitem[{\citenamefont{Wu et~al.}(2010)\citenamefont{Wu, Jiang, and
  Weng}}]{WuReview}
\bibinfo{author}{\bibfnamefont{M.~W.} \bibnamefont{Wu}},
  \bibinfo{author}{\bibfnamefont{J.~H.} \bibnamefont{Jiang}}, \bibnamefont{and}
  \bibinfo{author}{\bibfnamefont{M.~Q.} \bibnamefont{Weng}},
  \bibinfo{journal}{Phys. Rep.} \textbf{\bibinfo{volume}{493}},
  \bibinfo{pages}{61} (\bibinfo{year}{2010}).

\bibitem[{\citenamefont{Novoselov et~al.}(2004)\citenamefont{Novoselov, Geim,
  Morozov, Jiang, Zhang, Dubonos, Grigorieva, and Firsov}}]{NovoselovSCI2004}
\bibinfo{author}{\bibfnamefont{K.~S.} \bibnamefont{Novoselov}},
  \bibinfo{author}{\bibfnamefont{A.~K.} \bibnamefont{Geim}},
  \bibinfo{author}{\bibfnamefont{S.~V.} \bibnamefont{Morozov}},
  \bibinfo{author}{\bibfnamefont{D.}~\bibnamefont{Jiang}},
  \bibinfo{author}{\bibfnamefont{Y.}~\bibnamefont{Zhang}},
  \bibinfo{author}{\bibfnamefont{S.~V.} \bibnamefont{Dubonos}},
  \bibinfo{author}{\bibfnamefont{I.~V.} \bibnamefont{Grigorieva}},
  \bibnamefont{and} \bibinfo{author}{\bibfnamefont{A.~A.}
  \bibnamefont{Firsov}}, \bibinfo{journal}{Science}
  \textbf{\bibinfo{volume}{306}}, \bibinfo{pages}{666} (\bibinfo{year}{2004}).

\bibitem[{\citenamefont{Tombros et~al.}(2007)\citenamefont{Tombros, J\'{o}zsa,
  Popinciuc, Jonkman, and van Wees}}]{TombrosNat2007}
\bibinfo{author}{\bibfnamefont{N.}~\bibnamefont{Tombros}},
  \bibinfo{author}{\bibfnamefont{C.}~\bibnamefont{J\'{o}zsa}},
  \bibinfo{author}{\bibfnamefont{M.}~\bibnamefont{Popinciuc}},
  \bibinfo{author}{\bibfnamefont{H.~T.} \bibnamefont{Jonkman}},
  \bibnamefont{and} \bibinfo{author}{\bibfnamefont{B.~J.} \bibnamefont{van
  Wees}}, \bibinfo{journal}{Nature} \textbf{\bibinfo{volume}{448}},
  \bibinfo{pages}{571} (\bibinfo{year}{2007}).

\bibitem[{\citenamefont{Han et~al.}(2010)\citenamefont{Han, Pi, McCreary, Li,
  Wong, Swartz, and Kawakami}}]{KawakamiPRL2010}
\bibinfo{author}{\bibfnamefont{W.}~\bibnamefont{Han}},
  \bibinfo{author}{\bibfnamefont{K.}~\bibnamefont{Pi}},
  \bibinfo{author}{\bibfnamefont{K.~M.} \bibnamefont{McCreary}},
  \bibinfo{author}{\bibfnamefont{Y.}~\bibnamefont{Li}},
  \bibinfo{author}{\bibfnamefont{J.~J.~I.} \bibnamefont{Wong}},
  \bibinfo{author}{\bibfnamefont{A.~G.} \bibnamefont{Swartz}},
  \bibnamefont{and} \bibinfo{author}{\bibfnamefont{R.~K.}
  \bibnamefont{Kawakami}}, \bibinfo{journal}{Phys. Rev. Lett.}
  \textbf{\bibinfo{volume}{105}}, \bibinfo{pages}{167202}
  (\bibinfo{year}{2010}).

\bibitem[{\citenamefont{Yang et~al.}(2011)\citenamefont{Yang, Balakrishnan,
  Volmer, Avsar, Jaiswal, Samm, Ali, Pachoud, Zeng, Popinciuc
  et~al.}}]{GuntherodtBilayer}
\bibinfo{author}{\bibfnamefont{T.-Y.} \bibnamefont{Yang}},
  \bibinfo{author}{\bibfnamefont{J.}~\bibnamefont{Balakrishnan}},
  \bibinfo{author}{\bibfnamefont{F.}~\bibnamefont{Volmer}},
  \bibinfo{author}{\bibfnamefont{A.}~\bibnamefont{Avsar}},
  \bibinfo{author}{\bibfnamefont{M.}~\bibnamefont{Jaiswal}},
  \bibinfo{author}{\bibfnamefont{J.}~\bibnamefont{Samm}},
  \bibinfo{author}{\bibfnamefont{S.~R.} \bibnamefont{Ali}},
  \bibinfo{author}{\bibfnamefont{A.}~\bibnamefont{Pachoud}},
  \bibinfo{author}{\bibfnamefont{M.}~\bibnamefont{Zeng}},
  \bibinfo{author}{\bibfnamefont{M.}~\bibnamefont{Popinciuc}},
  \bibnamefont{et~al.}, \bibinfo{journal}{Phys. Rev. Lett.}
  \textbf{\bibinfo{volume}{107}}, \bibinfo{pages}{047206}
  (\bibinfo{year}{2011}).

\bibitem[{\citenamefont{Elliott}(1954)}]{Elliott}
\bibinfo{author}{\bibfnamefont{R.~J.} \bibnamefont{Elliott}},
  \bibinfo{journal}{Phys. Rev.} \textbf{\bibinfo{volume}{96}},
  \bibinfo{pages}{266} (\bibinfo{year}{1954}).

\bibitem[{\citenamefont{Yafet}(1963)}]{yafet1963g}
\bibinfo{author}{\bibfnamefont{Y.}~\bibnamefont{Yafet}},
  \bibinfo{journal}{Solid State Physics} \textbf{\bibinfo{volume}{14}},
  \bibinfo{pages}{1} (\bibinfo{year}{1963}).

\bibitem[{\citenamefont{Dyakonov and Perel}(1972)}]{DyakonovPerelSPSS1972}
\bibinfo{author}{\bibfnamefont{M.}~\bibnamefont{Dyakonov}} \bibnamefont{and}
  \bibinfo{author}{\bibfnamefont{V.}~\bibnamefont{Perel}},
  \bibinfo{journal}{Soviet Physics Solid State, USSR}
  \textbf{\bibinfo{volume}{13}}, \bibinfo{pages}{3023} (\bibinfo{year}{1972}).

\bibitem[{\citenamefont{Pikus and Titkov}(1984)}]{PikusTitkov}
\bibinfo{author}{\bibfnamefont{G.~E.} \bibnamefont{Pikus}} \bibnamefont{and}
  \bibinfo{author}{\bibfnamefont{A.~N.} \bibnamefont{Titkov}},
  \emph{\bibinfo{title}{{Spin relaxation under optical orientation in
  semiconductors}}} (\bibinfo{publisher}{Elsevier, Amsterdam},
  \bibinfo{year}{1984}), pp. \bibinfo{pages}{73--131}.

\bibitem[{\citenamefont{Szolnoki
  et~al.}(2017{\natexlab{a}})\citenamefont{Szolnoki, D\'ora, Kiss, Fabian, and
  Simon}}]{SzolnokiPRB2017}
\bibinfo{author}{\bibfnamefont{L.}~\bibnamefont{Szolnoki}},
  \bibinfo{author}{\bibfnamefont{B.}~\bibnamefont{D\'ora}},
  \bibinfo{author}{\bibfnamefont{A.}~\bibnamefont{Kiss}},
  \bibinfo{author}{\bibfnamefont{J.}~\bibnamefont{Fabian}}, \bibnamefont{and}
  \bibinfo{author}{\bibfnamefont{F.}~\bibnamefont{Simon}},
  \bibinfo{journal}{Phys. Rev. B} \textbf{\bibinfo{volume}{96}},
  \bibinfo{pages}{245123} (\bibinfo{year}{2017}{\natexlab{a}}).

\bibitem[{\citenamefont{Krau\ss{} et~al.}(2008)\citenamefont{Krau\ss{},
  Aeschlimann, and Schneider}}]{Schneider}
\bibinfo{author}{\bibfnamefont{M.}~\bibnamefont{Krau\ss{}}},
  \bibinfo{author}{\bibfnamefont{M.}~\bibnamefont{Aeschlimann}},
  \bibnamefont{and} \bibinfo{author}{\bibfnamefont{H.~C.}
  \bibnamefont{Schneider}}, \bibinfo{journal}{Phys. Rev. Lett.}
  \textbf{\bibinfo{volume}{100}}, \bibinfo{pages}{256601}
  (\bibinfo{year}{2008}).

\bibitem[{\citenamefont{Zhang and Wu}(2012)}]{WuNJP2012}
\bibinfo{author}{\bibfnamefont{P.}~\bibnamefont{Zhang}} \bibnamefont{and}
  \bibinfo{author}{\bibfnamefont{M.}~\bibnamefont{Wu}}, \bibinfo{journal}{New
  J. Phys.} \textbf{\bibinfo{volume}{14}}, \bibinfo{pages}{033015}
  (\bibinfo{year}{2012}).

\bibitem[{\citenamefont{Restrepo and Windl}(2012)}]{RestrepoWindl}
\bibinfo{author}{\bibfnamefont{O.~D.} \bibnamefont{Restrepo}} \bibnamefont{and}
  \bibinfo{author}{\bibfnamefont{W.}~\bibnamefont{Windl}},
  \bibinfo{journal}{Phys. Rev. Lett.} \textbf{\bibinfo{volume}{109}},
  \bibinfo{pages}{166604} (\bibinfo{year}{2012}).

\bibitem[{\citenamefont{Yu and Wu}(2016)}]{WuYuPRB2016}
\bibinfo{author}{\bibfnamefont{T.}~\bibnamefont{Yu}} \bibnamefont{and}
  \bibinfo{author}{\bibfnamefont{M.~W.} \bibnamefont{Wu}},
  \bibinfo{journal}{Phys. Rev. B} \textbf{\bibinfo{volume}{93}},
  \bibinfo{pages}{045414} (\bibinfo{year}{2016}).

\bibitem[{\citenamefont{Wang and Wu}(2014)}]{Wu_TMDC}
\bibinfo{author}{\bibfnamefont{L.}~\bibnamefont{Wang}} \bibnamefont{and}
  \bibinfo{author}{\bibfnamefont{M.~W.} \bibnamefont{Wu}},
  \bibinfo{journal}{Phys. Rev. B} \textbf{\bibinfo{volume}{89}},
  \bibinfo{pages}{115302} (\bibinfo{year}{2014}).

\bibitem[{\citenamefont{Boross et~al.}(2013)\citenamefont{Boross, D\'ora, Kiss,
  and Simon}}]{BorossSciRep2013}
\bibinfo{author}{\bibfnamefont{P.}~\bibnamefont{Boross}},
  \bibinfo{author}{\bibfnamefont{B.}~\bibnamefont{D\'ora}},
  \bibinfo{author}{\bibfnamefont{A.}~\bibnamefont{Kiss}}, \bibnamefont{and}
  \bibinfo{author}{\bibfnamefont{F.}~\bibnamefont{Simon}},
  \bibinfo{journal}{Sci. Rep.} \textbf{\bibinfo{volume}{3}},
  \bibinfo{pages}{3233} (\bibinfo{year}{2013}).

\bibitem[{\citenamefont{Szolnoki
  et~al.}(2017{\natexlab{b}})\citenamefont{Szolnoki, Kiss, Dóra, and
  Simon}}]{SzolnokiSciRep2017}
\bibinfo{author}{\bibfnamefont{L.}~\bibnamefont{Szolnoki}},
  \bibinfo{author}{\bibfnamefont{A.}~\bibnamefont{Kiss}},
  \bibinfo{author}{\bibfnamefont{B.}~\bibnamefont{Dóra}}, \bibnamefont{and}
  \bibinfo{author}{\bibfnamefont{F.}~\bibnamefont{Simon}},
  \bibinfo{journal}{Sci. Rep.} \textbf{\bibinfo{volume}{7}}
  (\bibinfo{year}{2017}{\natexlab{b}}).

\bibitem[{\citenamefont{Goussev et~al.}(2012)\citenamefont{Goussev, Jalabert,
  Pastawski, and Wisniacki}}]{Loschmidt_Scholarpedia}
\bibinfo{author}{\bibfnamefont{A.}~\bibnamefont{Goussev}},
  \bibinfo{author}{\bibfnamefont{R.~A.} \bibnamefont{Jalabert}},
  \bibinfo{author}{\bibfnamefont{H.~M.} \bibnamefont{Pastawski}},
  \bibnamefont{and} \bibinfo{author}{\bibfnamefont{D.~A.}
  \bibnamefont{Wisniacki}}, \bibinfo{journal}{Scholarpedia}
  \textbf{\bibinfo{volume}{7}}, \bibinfo{pages}{11687} (\bibinfo{year}{2012}),
  \bibinfo{note}{revision \#127578}.

\bibitem[{\citenamefont{Hahn}(1950)}]{Hahn1950}
\bibinfo{author}{\bibfnamefont{E.~L.} \bibnamefont{Hahn}},
  \bibinfo{journal}{Phys. Rev.} \textbf{\bibinfo{volume}{80}},
  \bibinfo{pages}{580} (\bibinfo{year}{1950}).

\bibitem[{\citenamefont{Murch et~al.}(2013)\citenamefont{Murch, Weber, Macklin,
  and Siddiqi}}]{QTrajectory1}
\bibinfo{author}{\bibfnamefont{K.~W.} \bibnamefont{Murch}},
  \bibinfo{author}{\bibfnamefont{S.~J.} \bibnamefont{Weber}},
  \bibinfo{author}{\bibfnamefont{C.}~\bibnamefont{Macklin}}, \bibnamefont{and}
  \bibinfo{author}{\bibfnamefont{I.}~\bibnamefont{Siddiqi}},
  \bibinfo{journal}{Nature} \textbf{\bibinfo{volume}{502}},
  \bibinfo{pages}{211} (\bibinfo{year}{2013}).

\bibitem[{\citenamefont{Ficheux et~al.}(2018)\citenamefont{Ficheux, Jezouin,
  Leghtas, and Huard}}]{QTrajectory2}
\bibinfo{author}{\bibfnamefont{Q.}~\bibnamefont{Ficheux}},
  \bibinfo{author}{\bibfnamefont{S.}~\bibnamefont{Jezouin}},
  \bibinfo{author}{\bibfnamefont{Z.}~\bibnamefont{Leghtas}}, \bibnamefont{and}
  \bibinfo{author}{\bibfnamefont{B.}~\bibnamefont{Huard}},
  \bibinfo{journal}{Nat. Comm.} \textbf{\bibinfo{volume}{9}}
  (\bibinfo{year}{2018}).

\bibitem[{\citenamefont{Minev et~al.}(2019)\citenamefont{Minev, Mundhada,
  Shankar, Gutiérrez-Jáuregui, Schoelkopf, Mirrahimi, Carmichael, and
  Devoret}}]{QTrajectory3}
\bibinfo{author}{\bibfnamefont{Z.~K.} \bibnamefont{Minev}},
  \bibinfo{author}{\bibfnamefont{S.~O.} \bibnamefont{Mundhada}},
  \bibinfo{author}{\bibfnamefont{P.}~\bibnamefont{Shankar},
  \bibfnamefont{S.and~Reinhold}},
  \bibinfo{author}{\bibfnamefont{R.}~\bibnamefont{Gutiérrez-Jáuregui}},
  \bibinfo{author}{\bibfnamefont{R.~J.} \bibnamefont{Schoelkopf}},
  \bibinfo{author}{\bibfnamefont{M.}~\bibnamefont{Mirrahimi}},
  \bibinfo{author}{\bibfnamefont{H.~J.} \bibnamefont{Carmichael}},
  \bibnamefont{and} \bibinfo{author}{\bibfnamefont{M.~H.}
  \bibnamefont{Devoret}}, \bibinfo{journal}{Nature}
  \textbf{\bibinfo{volume}{570}}, \bibinfo{pages}{200} (\bibinfo{year}{2019}).

\bibitem[{\citenamefont{Fabian et~al.}(2007)\citenamefont{Fabian,
  Matos-Abiaguea, Ertlera, Stano, and Zutic}}]{FabianActaPhysSlovaca}
\bibinfo{author}{\bibfnamefont{J.}~\bibnamefont{Fabian}},
  \bibinfo{author}{\bibfnamefont{A.}~\bibnamefont{Matos-Abiaguea}},
  \bibinfo{author}{\bibfnamefont{C.}~\bibnamefont{Ertlera}},
  \bibinfo{author}{\bibfnamefont{P.}~\bibnamefont{Stano}}, \bibnamefont{and}
  \bibinfo{author}{\bibfnamefont{I.}~\bibnamefont{Zutic}},
  \bibinfo{journal}{Acta Physica Slovaca} \textbf{\bibinfo{volume}{57}}
  (\bibinfo{year}{2007}).

\bibitem[{\citenamefont{M{\o}lmer et~al.}(1993)\citenamefont{M{\o}lmer, Castin,
  and Dalibard}}]{MCWF3}
\bibinfo{author}{\bibfnamefont{K.}~\bibnamefont{M{\o}lmer}},
  \bibinfo{author}{\bibfnamefont{Y.}~\bibnamefont{Castin}}, \bibnamefont{and}
  \bibinfo{author}{\bibfnamefont{J.}~\bibnamefont{Dalibard}},
  \bibinfo{journal}{J. Opt. Soc. Am. B} \textbf{\bibinfo{volume}{10}},
  \bibinfo{pages}{524} (\bibinfo{year}{1993}).

\bibitem[{\citenamefont{Dugaev et~al.}(2011)\citenamefont{Dugaev, Sherman, and
  Barna\ifmmode~\acute{s}\else \'{s}\fi{}}}]{ShermanPRB2011}
\bibinfo{author}{\bibfnamefont{V.~K.} \bibnamefont{Dugaev}},
  \bibinfo{author}{\bibfnamefont{E.~Y.} \bibnamefont{Sherman}},
  \bibnamefont{and}
  \bibinfo{author}{\bibfnamefont{J.}~\bibnamefont{Barna\ifmmode~\acute{s}\else
  \'{s}\fi{}}}, \bibinfo{journal}{Phys. Rev. B} \textbf{\bibinfo{volume}{83}},
  \bibinfo{pages}{085306} (\bibinfo{year}{2011}).

\bibitem[{\citenamefont{Kud\l{}a et~al.}(2018)\citenamefont{Kud\l{}a,
  Dyrda\l{}, Dugaev, Sherman, and Barna\ifmmode~\acute{s}\else
  \'{s}\fi{}}}]{ShermanPRB2018}
\bibinfo{author}{\bibfnamefont{S.}~\bibnamefont{Kud\l{}a}},
  \bibinfo{author}{\bibfnamefont{A.}~\bibnamefont{Dyrda\l{}}},
  \bibinfo{author}{\bibfnamefont{V.~K.} \bibnamefont{Dugaev}},
  \bibinfo{author}{\bibfnamefont{E.~Y.} \bibnamefont{Sherman}},
  \bibnamefont{and}
  \bibinfo{author}{\bibfnamefont{J.}~\bibnamefont{Barna\ifmmode~\acute{s}\else
  \'{s}\fi{}}}, \bibinfo{journal}{Phys. Rev. B} \textbf{\bibinfo{volume}{97}},
  \bibinfo{pages}{245307} (\bibinfo{year}{2018}).

\bibitem[{\citenamefont{Slichter}(1989)}]{SlichterBook}
\bibinfo{author}{\bibfnamefont{C.~P.} \bibnamefont{Slichter}},
  \emph{\bibinfo{title}{Principles of Magnetic Resonance}}
  (\bibinfo{publisher}{Spinger-Verlag}, \bibinfo{address}{New York},
  \bibinfo{year}{1989}), \bibinfo{edition}{3rd} ed.

\bibitem[{\citenamefont{Abragam}(1961)}]{AbragamBook}
\bibinfo{author}{\bibfnamefont{A.}~\bibnamefont{Abragam}},
  \emph{\bibinfo{title}{Principles of Nuclear Magnetism}}
  (\bibinfo{publisher}{Oxford University Press}, \bibinfo{address}{Oxford,
  England}, \bibinfo{year}{1961}).

\bibitem[{Sup()}]{SupMat}
\bibinfo{howpublished}{See Supplemental Material at [URL will be inserted by
  publisher] for additional discussion on the spin Boltzmann equation and the
  Loschmidt echo.}

\bibitem[{\citenamefont{Burkov and Balents}(2004)}]{BurkovBalents}
\bibinfo{author}{\bibfnamefont{A.~A.} \bibnamefont{Burkov}} \bibnamefont{and}
  \bibinfo{author}{\bibfnamefont{L.}~\bibnamefont{Balents}},
  \bibinfo{journal}{Phys. Rev. B} \textbf{\bibinfo{volume}{69}},
  \bibinfo{pages}{245312} (\bibinfo{year}{2004}).

\bibitem[{\citenamefont{Iordanskii et~al.}(1994)\citenamefont{Iordanskii,
  Lyanda-Geller, and Pikus}}]{Iordanskii_original}
\bibinfo{author}{\bibfnamefont{S.~V.} \bibnamefont{Iordanskii}},
  \bibinfo{author}{\bibfnamefont{Y.~B.} \bibnamefont{Lyanda-Geller}},
  \bibnamefont{and} \bibinfo{author}{\bibfnamefont{G.~E.} \bibnamefont{Pikus}},
  \bibinfo{journal}{JETP} \textbf{\bibinfo{volume}{60}}, \bibinfo{pages}{206}
  (\bibinfo{year}{1994}).

\bibitem[{\citenamefont{Knap et~al.}(1996)\citenamefont{Knap, Skierbiszewski,
  Zduniak, Litwin-Staszewska, Bertho, Kobbi, Robert, Pikus, Pikus, Iordanskii
  et~al.}}]{Iordanskii}
\bibinfo{author}{\bibfnamefont{W.}~\bibnamefont{Knap}},
  \bibinfo{author}{\bibfnamefont{C.}~\bibnamefont{Skierbiszewski}},
  \bibinfo{author}{\bibfnamefont{A.}~\bibnamefont{Zduniak}},
  \bibinfo{author}{\bibfnamefont{E.}~\bibnamefont{Litwin-Staszewska}},
  \bibinfo{author}{\bibfnamefont{D.}~\bibnamefont{Bertho}},
  \bibinfo{author}{\bibfnamefont{F.}~\bibnamefont{Kobbi}},
  \bibinfo{author}{\bibfnamefont{J.~L.} \bibnamefont{Robert}},
  \bibinfo{author}{\bibfnamefont{G.~E.} \bibnamefont{Pikus}},
  \bibinfo{author}{\bibfnamefont{F.~G.} \bibnamefont{Pikus}},
  \bibinfo{author}{\bibfnamefont{S.~V.} \bibnamefont{Iordanskii}},
  \bibnamefont{et~al.}, \bibinfo{journal}{Phys. Rev. B}
  \textbf{\bibinfo{volume}{53}}, \bibinfo{pages}{3912} (\bibinfo{year}{1996}).

\bibitem[{\citenamefont{Portis}(1953)}]{PortisPR1953}
\bibinfo{author}{\bibfnamefont{A.~M.} \bibnamefont{Portis}},
  \bibinfo{journal}{Phys. Rev.} \textbf{\bibinfo{volume}{91}},
  \bibinfo{pages}{1071} (\bibinfo{year}{1953}).

\end{thebibliography}

\end{widetext}
\newpage

%\end{document}

\end{document}